\documentclass{emulateapj}
\usepackage{apjfonts}
\newcommand{\scaleup}{\epsscale{1.1}}

\newcommand{\plotter}{\plotone}

\newcommand{\mdot}{\lambda}

\newcommand{\msun}{M_{\sun}}

\newcommand{\dtdlogl}{{\rm d}t/{\rm d}\log{L}}
\newcommand{\dtdlogL}{\dtdlogl}
\newcommand{\tq}{t_{Q}}

\shorttitle{Testing Quasar Lifetime Models}
\shortauthors{Hopkins \&\ Hernquist}
\slugcomment{Submitted to ApJ, September 8, 2008}
\begin{document}

\title{Quasars Are Not Light-Bulbs: Testing Models of Quasar 
Lifetimes with the Observed Eddington Ratio Distribution}
\author{Philip F. Hopkins\altaffilmark{1,2} 
\&\ Lars Hernquist\altaffilmark{1}
}
\altaffiltext{1}{Harvard-Smithsonian Center for Astrophysics, 
60 Garden Street, Cambridge, MA 02138}
\altaffiltext{2}{Department of Astronomy, University of California 
Berkeley, Berkeley, CA 94720} 

\begin{abstract}

We use the observed distribution of Eddington ratios as a function of supermassive 
black hole (BH) mass to constrain models of quasar/AGN lifetimes and lightcurves. 
Given the observed (well constrained) AGN luminosity function, a 
particular model 
for AGN lightcurves $L(t)$ or, equivalently, the distribution of AGN lifetimes 
(time above a given luminosity $t(>L)$) translates directly and 
uniquely (without further assumptions) to a predicted distribution of Eddington 
ratios at each BH mass. Models for self-regulated BH growth, 
in which feedback produces a self-regulating ``decay'' or ``blowout'' phase 
after the AGN reaches some peak luminosity/BH mass and begins to 
expel gas and shut down accretion, make specific predictions for 
the lightcurves/lifetimes, distinct from e.g.\ the expected distribution if 
AGN simply shut down by gas starvation (without feedback) and very different 
from the prediction of simple phenomenological ``light bulb'' 
scenarios. 
We show that the present observations of the Eddington ratio distribution, 
spanning nearly 5 orders of magnitude in Eddington ratio, 3 orders of 
magnitude in BH mass, and redshifts $z=0-1$, agree well with the 
predictions of self-regulated models, and rule out phenomenological 
``light bulb'' or pure exponential models, as well as gas starvation 
models, at high significance ($\sim5\,\sigma$). We also compare with 
observations of the  
distribution of Eddington ratios at a given AGN luminosity, and find similar 
good agreement (but show that these observations are much less constraining). 
We fit the functional form 
of the quasar lifetime distribution and provide 
these fits for use, and show how the Eddington ratio distributions 
place precise, tight limits on the AGN lifetimes at various luminosities, 
in agreement with model predictions.  We compare with  
independent estimates of episodic lifetimes and use this to 
constrain the shape of the typical AGN lightcurve, and 
provide simple analytic fits to these for use in other analyses. Given these 
constraints, the average local BH must have gained its mass in no more than a couple of 
bright, near peak-luminosity episodes, in agreement with models of 
accretion triggering in interactions and mergers. 

\end{abstract}


\section{Introduction}
\label{sec:intro}

Quasars and active galactic nuclei (AGN) 
are among the most luminous, energetic, and distant 
objects in the Universe. Simple integral arguments 
\citep{soltan82} make it clear that the supermassive black hole (BH) 
population was grown primarily through accretion in luminous 
AGN phases, and that the accretion luminosity released in these 
phases dominates the X-ray background and constitutes a large
fraction of the bolometric energy production of the Universe. 
Comparison of e.g.\ the clustering \citep{croom:clustering,porciani2004,
hopkins:clustering} and host galaxy properties 
\citep{bahcall:qso.hosts,canalizostockton01:postsb.qso.mergers,
dunlop:qso.hosts,hopkins:red.galaxies,zakamska:qso.hosts,
zakamska:mir.seds.type2.qso.transition.at.special.lum} of 
high and low redshift AGN and galaxies demonstrates that 
AGN are the progenitors of modern-day spheroids.
 
Moreover, with the discovery of tight correlations 
between the masses of black holes and the velocity dispersion 
\citep{FM00,Gebhardt00}, masses \citep{magorrian}, and 
perhaps most fundamentally binding energy or potential 
well depth \citep{hopkins:bhfp.obs,hopkins:bhfp.theory,aller:mbh.esph} 
of the host demonstrates a 
fundamental link between the growth of supermassive black holes 
and galaxy formation. A number of models have been developed
arguing that the energy or momentum released 
from an accreting supermassive black hole, even if only a small fraction 
couples to the surrounding ISM, is sufficient to 
halt further accretion onto the black hole and drive away gas, 
self-regulating growth by shutting off 
the quasar\footnote{In what follows, we use the term ``quasar'' somewhat loosely, 
as a proxy for high-Eddington ratio accretion activity, rather than as a reference 
to specific optical properties. We use the term AGN to refer to BH 
accretion at all levels.} 
and quenching star formation in 
the galaxy and therefore allowing it to redden rapidly 
\citep[see e.g.][]{ciottiostriker:cooling.flow.selfreg.1,
ciottiostriker:cooling.flow.selfreg.2,silkrees:msigma,burkertsilk:msigma,
dimatteo:msigma,hopkins:lifetimes.letter,hopkins:qso.all,
murray:momentum.winds,sazonov:radiative.feedback, springel:models,springel:red.galaxies}. 

One of the most basic aspects of black hole growth, and a
powerful test of these self-regulating models for AGN evolution, is the
quasar/AGN lifetime. Given a sufficiently well-known 
lifetime distribution (lifetime as a function of 
e.g.\ luminosity, black hole mass, and other properties), the 
well-constrained quasar luminosity function (QLF) can be 
empirically translated (without invoking any models or 
additional assumptions) into the triggering rate of AGN as a function of e.g.\ 
luminosity, BH/host galaxy/dark matter halo mass, redshift, and other properties, 
as well as the active BH mass function, Eddington ratio and 
duty cycle distributions, and differential (mass-dependent) rate of buildup of 
the BH mass function. 

Observations generally constrain {\em quasar} lifetimes to
the range $\approx 10^{7}-10^{8}$ yr \citep[for a review,
see][]{martini04}.  These estimates are primarily based on demographic
or integral arguments which combine observations of the present-day
population of supermassive black holes and accretion by the
high-redshift quasar population
\citep[e.g.,][]{soltan82,haehnelt:bh.synthesis.model,yutremaine:bhmf,yulu:bhmf,
haiman:bhmf,marconi:bhmf,shankar:bhmf}, or incorporate quasars into
models of galaxy evolution \citep[e.g.,][]{kh00,wyitheloeb:sam,
dimatteo:cosmo.bh.growth.sim.1,dimatteo:qso.host.metals,
dimatteo:cosmo.bhs,granato:sam,scannapieco:sam,lapi:qlf.sam,
hopkins:qso.all,hopkins:groups.qso,sijacki:radio} 
or reionization of HeII \citep{sokasian:heII.reion.epoch,
sokasian:heII.reion,faucher:heII.reion.qso.constraints,faucher:ion.background.evol,
mcquinn:helium.reionization.model}. 
Results from clustering in quasar surveys
\citep[e.g.,][]{porciani2004,grazian:local.qso.clustering,
croom:clustering,myers:clustering.old,myers:clustering,
lidz:clustering,porciani:clustering,shen:clustering,
daangela:clustering,hopkins:clustering}, the proximity effect in the
Ly$\alpha$ forest \citep{bajtlik:gunnpetersen.qso.age.est,
haiman:gunnpetersen.qso.age.est,yulu:stromgren,faucher:proximity}
\citep[but see also][]{lidz:proximity}, 
and the transverse proximity effect 
\citep{jakobsen:heII.ion.transverse.proximity,schirber:transverse.proximity,
worseck06:indirect.transverse.proximity,worseck07:indirect.transverse.proximity,
goncalves:transverse.proximity}  similarly suggest
lifetimes $\sim{\rm a\ few}\,\times 10^{7}\,$yr.

These observations, however, pertain in particular to the {\em quasar} lifetime -- 
i.e.\ the characteristic time spent at high Eddington ratios/accretion rates, 
where much of the mass of a BH is accreted. Unsurprisingly, 
the observations suggest a lifetime similar to the \citet{salpeter64} 
time (the $e$-folding time for Eddington-limited black hole growth) 
$t_{S}=4.2\times10^{7}\,(\epsilon_{r}/0.1)\,$yr for accretion
with radiative efficiency $\epsilon_{r}=L/\dot{M}_{\rm BH} c^{2}\sim0.1$. 
AGN, however, are not a homogeneous population, and so 
the lifetime is not a single number -- in general, 
it should be a function of luminosity and other parameters such as 
BH or host mass and (possibly) various physical effects. As 
advocated by \citet{hopkins:lifetimes.letter,hopkins:lifetimes.methods}, 
the AGN lifetime should properly be thought of as a 
{\em luminosity-dependent lifetime} distribution -- i.e.\ some 
time $t(>L)$ as a function of $L$ or differential $\dtdlogL$. 

\citet{hopkins:lifetimes.letter,hopkins:lifetimes.methods,
hopkins:qso.all,hopkins:faint.slope,hopkins:seyferts} study this luminosity-dependent 
lifetime/duty cycle distribution in hydrodynamic simulations and analytic models 
of feedback-regulated BH growth, and show that such self-regulation 
leads to a generic and unique predicted form for the lifetime distribution. 
After some initial trigger that fuels gas inflows 
\citep[such as e.g.\ major and minor mergers or disk instabilities; 
for discussion see][]{dimatteo:msigma,hopkins:seyferts,
hopkins:cusps.mergers,hopkins:cusps.ell,hopkins:cores,hopkins:cusps.fp,
hopkins:cusps.evol,hopkins:disk.survival,hopkins:seyfert.limits,younger:minor.mergers}, 
AGN grow in approximately Eddington-limited fashion
until reaching some critical mass where, if some small fraction of the radiant 
energy/momentum can couple to the surrounding ISM, the feedback is sufficient to halt 
inflows and/or expel gas and shut down future accretion. This ``upper limit'' to 
growth is essentially an Eddington limit effective at the scales where the 
host galaxy, rather than the BH, dominates the gravitational potential, and 
therefore is set not by the details of fueling mechanisms, but instead
by (relatively generic) global parameters of the accretion physics
such as the host galaxy mass and AGN luminosity. 

That the resulting lifetime distribution is independent of fueling mechanism 
(i.e.\ is not specifically related to, for example, merger-induced fueling, but 
is a generic consequence of models where BH growth is self-regulated by feedback) 
has been demonstrated in the 
nearly identical lifetime distributions obtained by models of fueling in 
major mergers, minor mergers, flyby events, 
bar instabilities, and random ``stochastic'' encounters with 
nuclear molecular clouds, under similar feedback-regulated 
conditions \citep{hopkins:qso.all,hopkins:faint.slope,hopkins:seyferts,
johansson:mixed.morph.mbh.sims,
younger:minor.mergers}. These various fueling mechanisms do result in 
other important differences (in e.g.\ host properties and spectral properties of 
observed AGN), and they will lead to different evolution of BHs in a cosmological 
sense (global triggering rates and their evolution varying significantly
for the mechanisms above) and produce BHs of different masses 
(to the extent that they produce bulges across a large range
of masses, their self-regulating
nature and the existence of the BH-host correlations ensures this to be
the case). 
These are discussed in more detail in other papers \citep[see e.g.][]{hopkins:seyfert.limits}, 
but the important point is that, for a given triggering rate at some redshift 
and BH mass interval (set by some cosmological or galactic processes), 
self-regulated models predict a similar 
effective Eddington limit and lightcurve.

Just as growth at the 
traditional Eddington limit leads to a self-similar solution for the AGN lightcurve -- 
exponential growth -- the expulsion of gas in this analogous limit leads to 
a self-similar lightcurve once the gas begins to be removed from 
the vicinity of the BH -- a power-law decay of the form 
$L\propto t^{-(1.5-2.0)}$. In turn, this translates into a lifetime distribution 
$\dtdlogL$ with a characteristic faint-end (low-$L$) power-law like behavior: 
i.e.\ the time spent above a given luminosity/Eddington ratio (at fixed final 
BH mass) scales $\propto L^{-\beta}$ with a $\beta\sim0.6$ at low-$L$, with 
a cutoff near the Eddington limit/peak luminosity of the system. 
The normalization is set by the characteristic timescales of the system -- 
the Salpeter time and the (very similar) characteristic dynamical times 
in the central regions of the galaxy -- naturally yielding a robust prediction of the 
observed quasar lifetime as well as the complete AGN lifetime distribution. 

This is significantly different from what is assumed in commonly adopted 
phenomenological models, as well as other physical prescriptions. Quasars 
are often treated crudely
as ``light-bulbs'' -- i.e.\ assumed to be ``on'' for a 
fixed time (the ``quasar lifetime'') at fixed luminosity or Eddington ratio, 
and otherwise ``off.'' In 
such a case, the lifetime/Eddington ratio distribution does not 
increase towards lower luminosities, but instead is strongly peaked 
about a characteristic high Eddington ratio/luminosity (strictly speaking, a 
$\delta$-function; or in a Schechter-function parameterization with some 
scatter, $\beta\ll 0$). Similar results are obtained if one assumes 
that quasar 
lightcurves are pure exponentials (corresponding to growth at 
fixed Eddington ratio with either an instantaneous
cutoff or time-mirrored 
exponential luminosity decay), which yields equal time spent per 
logarithmic interval in luminosity (i.e.\ a luminosity-independent lifetime, 
or $\beta=0$ Schechter function). 

More physically motivated but distinct models 
make their own predictions for the lifetime distribution.  For example, 
if one assumes that the quasar accretion is regulated by a standard 
\citet{shakurasunyaev73} thin disk, and the fuel supply is immediately
removed but there is no feedback (i.e.\ a gas starvation scenario), 
one obtains a similarity solution for the accretion rate versus time 
that yields a lifetime distribution more akin to, but still significantly 
distinct from that predicted in self-regulated models \citep[see e.g.][]{yu:mdot.dist}. 

A number of indirect tests have been proposed to break the degeneracies 
between these models and constrain the quasar lifetime distribution, 
and the observations in these scenarios have thus far supported 
the predictions of self-regulated models. 
These include: the dependence of quasar clustering on 
luminosity \citep{lidz:clustering,hopkins:clustering}, where observations finding 
e.g.\ a weak dependence of clustering amplitude on luminosity (at fixed 
redshift) support lifetime models with more time spent at lower $L$ 
\citep{adelbergersteidel:lifetimes,coil:agn.clustering,
myers:clustering,daangela:clustering}; the shape of the active BH mass function in various 
luminosity-selected samples \citep{hopkins:lifetimes.interp,
hopkins:qso.all,hopkins:groups.qso}, 
including more massive systems at lower $L$ (in extreme cases even being 
peaked) rather than tracing an identical shape to the QLF 
\citep{heckman:local.mbh,greene:active.mf}; the evolution in the faint-end QLF 
slope with redshift \citep{hopkins:faint.slope}, flattening (weakly) with 
redshift \citep{ueda03:qlf,hasinger05:qlf,lafranca:hx.qlf,silverman:hx.spacedensity.ldde,
hopkins:bol.qlf,fontanot:highz.qlf,silverman:hx.lf,siana:z3.swire.qlf}
as predicted in some self-regulated models owing to a (weak) dependence of 
quasar lifetime distributions on BH mass/peak luminosity; the 
shape of the distribution in quasar host galaxy masses/luminosities 
\citep{hamilton:qso.host.lf,hopkins:merger.lfs}, similar in nature to the active 
BH mass function as a test of lifetime models; the mass functions and 
clustering as a function of mass of quasar ``remnants'' 
\citep[i.e.\ bulges/spheroids, 
given the observed $M_{\rm BH}-\sigma$ relation; see][]{hopkins:red.galaxies,
hopkins:groups.ell,hopkins:transition.mass,bundy:mtrans,haiman:qlf.from.ell.ages,
shankar:implied.msig.from.gal.ages,yulu:lightcurve.constraints.from.bhmf.integration}; 
and the relation between 
observed luminosity functions in different bands and active BH 
masses \citep{hopkins:lifetimes.obscuration,shankar:bol.qlf,
merloni:synthesis.model,bundy:agn.lf.to.mf.evol,yulu:lightcurve.constraints.from.bhmf.integration}. 

Although these observations are consistent with the predictions from 
self-regulated models, they are indirect, and as such are not able to 
rule out alternative interpretations (in the case of e.g.\ clustering or 
the evolution of the faint-end QLF slope) 
or depend on additional (albeit observationally and physically-motivated) 
assumptions. Moreover, many are restricted to relatively bright Seyfert/quasar 
populations, where the models are all similar (near these luminosities, 
they all predict that the entire population must be dominated by relatively massive 
BHs at high Eddington ratios $\sim 0.1-1$, and given the rapid growth 
at these Eddington ratios, the lifetime in this regime must be similar, comparable to 
the \citet{salpeter64} time). Biases introduced (selecting for specific Eddington ratios, 
for example) will also be important in samples selected by e.g.\ broad emission lines 
or optical/UV/IR colors/spectral shape \citep[see][]{hopkins:seyfert.bimodality}. 

The observed Eddington ratio distributions -- specifically, the complete 
distribution of Eddington ratios for {\em all} BHs of a given mass (obscured 
or unobscured, luminous or under-luminous), however, represent a direct and 
powerful test of these models, with the predicted behavior at lower Eddington ratios/luminosities 
being strongly divergent. Given the QLF, a lifetime model directly and uniquely 
translates (without any additional assumptions) into a distribution of 
Eddington ratios at each BH mass, and vice versa. The requirements 
are demanding: breaking these degeneracies necessitates large samples, complete to 
all objects of a given BH mass, in a large volume (to constrain rare high-Eddington ratio 
objects), but sufficiently deep to measure even faint levels of AGN activity in those 
objects. Furthermore, a probe of AGN activity such as X-ray or narrow-line emission, 
robust to obscuration effects (or the possible disappearance of the broad-emission line 
region and/or thin disk at low Eddington ratios) is important both to obtain a complete 
census of AGN activity and to avoid biases from e.g.\ luminosity or Eddington ratio-dependent 
obscuration/dilution/SED shapes. Fortunately, with the advent of wide area 
spectroscopic surveys such as the SDSS, it has become possible to constrain the 
Eddington ratio distribution at low redshifts, over a sufficiently large dynamic 
range to break the degeneracies between these models, as a function of 
various AGN and galaxy properties. 

Here, we combine a large number of observations of AGN Eddington ratio 
distributions as a function of BH and host galaxy mass, AGN luminosity, and 
redshift, in order to test these models and tightly constrain critical quantities 
such as the quasar lifetime as a function of luminosity, and show how the 
present observations are already sufficient to rule out, at high significance, a 
wide variety of alternative, simplified physical and phenomenological AGN 
lifetime/lightcurve models. 

In \S~\ref{sec:compare.dist} we compare these model predictions with observations 
of the Eddington ratio distribution measured directly at $z=0$ over a range of 
BH mass, and inferred indirectly at $z=0-1$. In \S~\ref{sec:compare.L} we 
similarly compare with Eddington ratio distributions measured not at fixed BH mass, 
but at fixed AGN luminosity, again over the range $z=0-1$. 
In \S~\ref{sec:compare.models} we show how these observations tightly constrain 
physical and phenomenological models for the quasar lifetime/accretion 
rate distribution, relate to a possible dependence of quasar feedback 
efficiency on mass, and rule out a number of alternative lifetime/lightcurve models. We 
show how the observations tightly constrain even general, parameterized lifetime 
models to a narrow range about the physical models, and can directly be converted 
to yield the cosmologically integrated AGN lifetime and $z=0$ duty cycles 
as a function of Eddington ratio. In \S~\ref{sec:lightcurves} we translate these 
Eddington ratio/lifetime distribution constraints to limits on the form of the 
``typical'' AGN lightcurve, and discuss constraints on the ``episodic'' quasar lifetime 
and how, combined with the duty cycle constraints, this can give a bound on the 
number of accretion episodes per AGN and shape of the typical lightcurve. 
In \S~\ref{sec:implications} we demonstrate the constraints from these observations on how 
much mass present-day BHs accreted in various intervals in Eddington ratio and 
luminosity. 
Finally, in \S~\ref{sec:discussion} we discuss our results, 
the implications of the observations for a broad range of 
AGN properties, and the prospects for 
future observational tests. 

For ease of comparison, we convert all observations 
to bolometric luminosities given the appropriate bolometric corrections 
from \citet{hopkins:bol.qlf} \citep[see also][]{elvis:atlas,richards:seds}. 
We adopt a $\Omega_{\rm M}=0.3$, $\Omega_{\Lambda}=0.7$,
$H_{0}=70\,{\rm km\,s^{-1}\,Mpc^{-1}}$ cosmology and normalize all 
observations and models appropriately \citep[note that this generally affects only 
the exact normalization of quantities here, not the qualitative conclusions, 
and differences are negligible within the range of cosmologies allowed by 
present constraints; e.g.][]{komatsu:wmap5}.

\section{Comparing With Complete Eddington Ratio Distributions} 
\label{sec:compare.dist}

Given the QLF $\Phi(L\,|\,z)$ and some model for the quasar 
lifetime or differential time at different Eddington ratios: 
\begin{equation}
\frac{{\rm d}t}{{\rm d}\log{L}}(L\,|\,M_{\rm BH}) \equiv 
\frac{{\rm d}}{{\rm d}\log{L}}{\Bigl [} t(L^{\prime} > L\, |\, M_{\rm BH}) {\Bigr ]}, 
\label{eqn:dtdlogl.defn}
\end{equation}
it is straightforward to de-convolve and determine the Eddington ratio 
distribution. For example, if quasars were ``light bulbs'' that shine either 
at some fixed high Eddington ratio 
\begin{equation}
\mdot\equiv \frac{L}{L_{\rm Edd}} = \frac{L_{\rm bol}}{1.3\times10^{38}\,{\rm erg\,s^{-1}}\,
(M_{\rm BH}/\msun)}
\label{eqn:lambda.defn}
\end{equation}
with $\mdot_{\rm on}\sim1$ in an ``on'' state for a time 
$\tq$ and $\mdot\ll1$ in an ``off'' state at other times, then 
the implied lifetime distribution $\dtdlogl$ is a delta-function at 
$\mdot_{\rm on}$. The observed QLF in such a case is, of course, a linear 
translation of the active BH mass function (with $\mdot_{\rm on}$ determining the 
re-normalization or shift in the horizontal axis, from $M_{\rm BH}$ to $L$, and 
the absolute value of the quasar lifetime $\tq$ determining the abundance/vertical 
axis). So, at a given $M_{\rm BH}$, the Eddington ratio distribution will 
be a delta-function at $\mdot_{\rm on}$ with a normalization/fractional 
abundance at this $\mdot_{\rm on}$ determined by the number of active 
quasars (of the corresponding luminosity) at the observed redshift. 

In general, in fact, so long as the quasar lifetime at a given $L$ is short 
compared to the Hubble time (i.e.\ cosmological evolution in e.g.\ triggering 
rates can be ignored in the constraint), then the Eddington ratio distribution 
at a given $M_{\rm BH}$ will be $\dtdlogl$ (modulo a normalization reflecting 
the ``on'' fraction or $\tq/t_{H}$, where 
$t_{H}$ is the Hubble time at the given redshift) -- i.e.\ we can trivially 
translate to a ``duty cycle'' distribution (fractional population at each $L$ or 
$\mdot$):\footnote{The translation between the distribution in $\log L$ and $\log{\lambda}$ is trivial 
in terms of observations at a given BH mass; in terms of converting model predictions 
between one and the other, although the relation is not completely trivial, we find 
in practice that the two are nearly equivalent, especially at Eddington ratios $\lesssim0.2$ where 
most of the data with which we will compare lie, since some initial 
$M_{\rm BH}$ does not change much as a model system moves through a low Eddington ratio 
phase. We will therefore use the two interchangeably in this paper as is proper for the 
observations, but have converted all physical models to the appropriate representation.}
\begin{equation}
\frac{{\rm d}\delta}{{\rm d}\log{\lambda}}=
\frac{{\rm d}\delta}{{\rm d}\log{L}} \approx \frac{1}{t_{H}(z)}\,\frac{{\rm d}t}{{\rm d}\log{L}}\ \ \ \ (t\ll t_{H}).
\label{eqn:duty.cycle.translation}
\end{equation}
By definition of the duty cycle, of course, this directly relates to the actual number density function of 
BHs at a given mass and luminosity or accretion rate 
\begin{equation}
\Phi(\lambda\,|\,M_{\rm BH}) \equiv \frac{{\rm d}n(\lambda,\,M_{\rm BH})}{{\rm d}\log{\lambda}}
= n(M_{\rm BH})\,\frac{{\rm d}\delta}{{\rm d}\log{\lambda}}\ . 
\label{eqn:numberdensity}
\end{equation}
The shape of the observed Eddington 
ratio distribution, therefore, contains information about the shape of 
$\dtdlogL$ independent of either normalization, and vice versa. Such observations 
hence provide a useful and direct probe of quasar accretion rate 
distributions. 

If the AGN population (e.g.\ triggering rates or number of bright objects) 
is relatively constant for systems of a given mass over the redshift range of interest (i.e.\ if 
the population is still growing, at least in a statistical sense, around the observed redshifts) 
and the lifetime is short relative to the Hubble time, then Equation~\ref{eqn:duty.cycle.translation} 
is applicable, and the observed Eddington ratio distribution, independent of 
any other constraints, can be directly translated to the lifetime distribution. 
When the triggering rate evolves strongly with redshift and/or the characteristic lifetime 
is long compared to the Hubble time (e.g.\ for massive systems at low redshift, where 
their growth is dominated by higher-redshift periods), the Eddington ratio distribution at a 
given $z$ is 
still uniquely predicted by a given $\dtdlogL$ model, but this must be convolved over 
the redshift distribution of activity (i.e.\ given the observed QLF and $\dtdlogL$, we integrate over 
time to determine the predicted ${\rm d}{\rm \delta}/{\rm d}\log{\lambda}$). Even this, in practice, 
does not significantly change the direct mapping between the shape of 
the observed Eddington ratio distribution and $\dtdlogl$; it mainly amounts to deriving 
a more correct effective ``duty cycle'' (normalization) than multiplying by $\sim1/t_{H}$.
For more discussion and details of the relevant equations, 
we refer to \citet{yulu:bhmf,hopkins:lifetimes.methods,
hopkins:qso.all,hopkins:faint.slope}; for now, we note that the 
observed QLF $\Phi(L\,|\,z)$ is well-constrained at all the luminosities and 
redshifts of interest (since the relevant observations are primarily at low 
redshifts $z\lesssim1$), and so any given model for $\dtdlogl$ 
{\em uniquely} translates to an Eddington 
ratio distribution as a function of BH mass or AGN luminosity.  We wish to
compare these distributions to the observations. 

In order to do so, we want to begin with a complete distribution of BH accretion 
rates at a given BH mass. It is important to do so -- in models where e.g.\ systems 
of a given mass can have a broad range of Eddington ratios, the 
distribution of Eddington ratios at a given BH mass can be {\em qualitatively} 
very different from the distribution at a given luminosity (see \S~\ref{sec:compare.L}). 
Moreover, if what is desired 
is a distribution at all possible Eddington ratios, then there is no useful definition 
of an ``active'' AGN: what we really desire is to begin with a complete census of all 
BHs of a given mass (active or inactive at any level), and to measure the Eddington ratio 
distribution within this sample. 

Fortunately, the existence of a tight correlation between 
host galaxy luminosity/stellar mass/velocity dispersion and BH mass 
makes this possible: \citet{heckman:local.mbh} and 
\citet{yu:mdot.dist} select complete samples of all SDSS galaxies with a 
narrow range in velocity dispersion $\sigma$ (and corresponding narrow range in 
BH mass), and then examine this sample to a limiting depth for narrow equivalent width 
AGN features \citep[following the methodology and 
classifications in][]{kauffmann:qso.hosts,kewley:agn.host.sf}. 
Given a bolometric correction, this allows a complete 
census of all activity at a given BH mass down to some well-known limit in $\mdot$. 

Moreover, the use of narrow lines is helpful for three reasons, as opposed to e.g.\ use of 
optical broad lines or an optical/IR color cut in identification of AGN and 
determination of their bolometric luminosities. 
First, it can probe very faint AGN and does not introduce much bias in terms of e.g.\ the risk of 
the host light in brighter systems diluting the AGN. Second, 
it allows us to include the obscured/Type 2 population \citep[the abundance of which 
may depend on luminosity; see e.g.][]{ueda03:qlf,
simpson:type1.frac,lafranca:hx.qlf,barger:qlf,beckmann:very.hx.qlf,bassani:integral.qlf,
gilli:obscured.fractions,hasinger:absorption.update}. 
Third, it should 
enable us to identify quasars even in states of moderate radiative inefficiency: there is 
growing evidence that AGN at low Eddington ratios may transition to a radiatively 
inefficient state characterized by the absence of a thin disk.  Such objects 
are still accreting but appear primarily as X-ray (because a hot corona survives 
this transition) and narrow-line sources, rather than broad-line or optical continuum sources 
\citep[see e.g.][]{NY94,nym95,nym96,meier:jets.in.adaf,maccarone:agn.riaf.connection,
xbongs,jester:riaf.test,mcclintock:xrb.review,cao:riaf.constraints}. 
We discuss these distinctions in more detail in a companion paper 
\citep{hopkins:seyfert.bimodality}, and 
outline how they can influence e.g.\ the Eddington ratio distribution determined 
via various selection criteria. For our purposes here, however, the data adopted 
either avoid these uncertainties owing to their selection/identification 
methodology, or cover a luminosity range where these concerns are not important 
\citep[see also][]{heckman:local.mbh,kauffmann:qso.hosts}. 

\begin{figure*}
    \centering
    \plotone{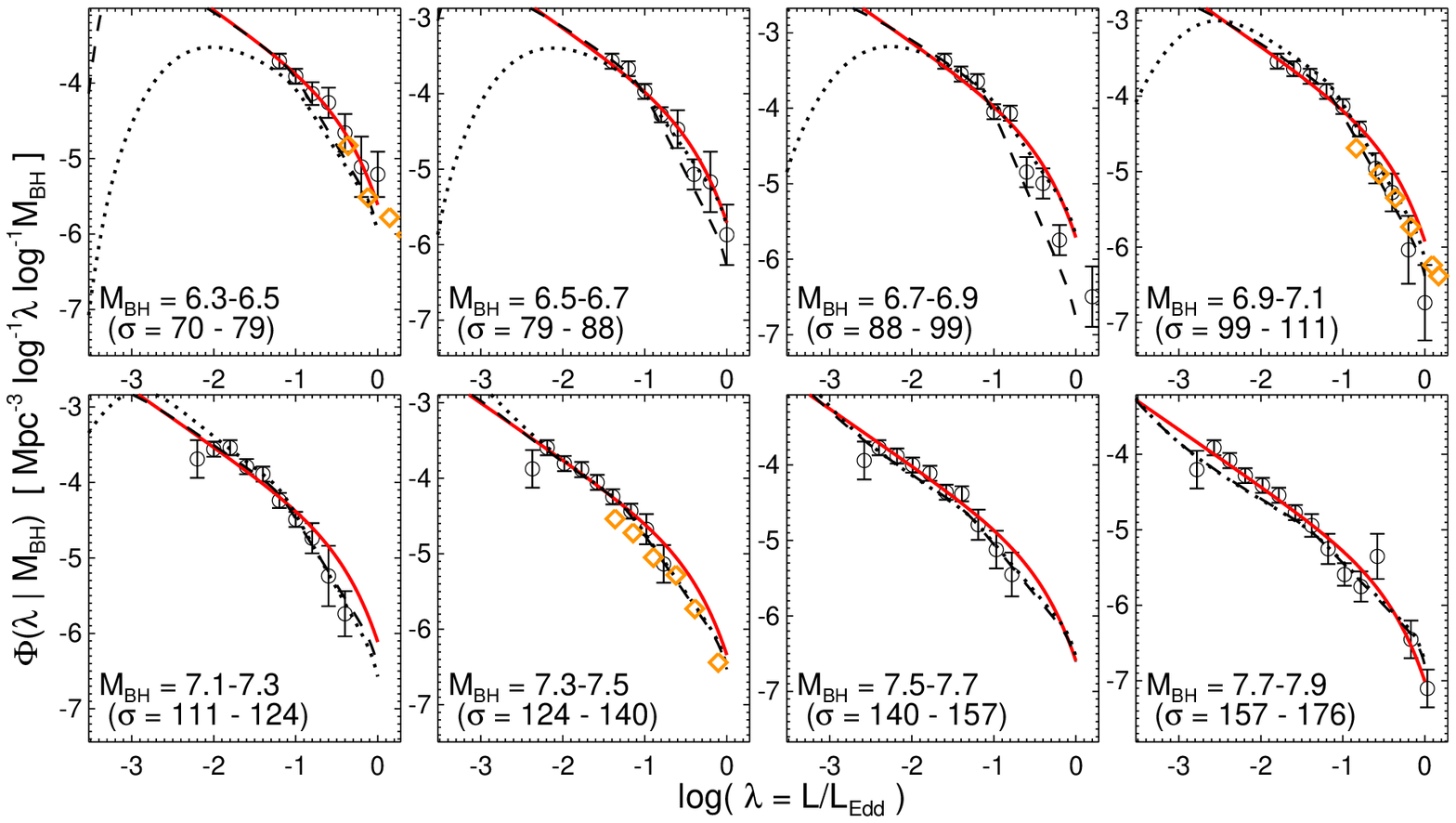}
    \caption{Distribution of Eddington ratios at a given BH mass (at $z<0.2$). 
    We compare the observed distribution from SDSS narrow-line objects 
    \citep[][black circles with error bars and orange diamonds, respectively]{yu:mdot.dist,
    heckman:local.mbh} 
    to the distribution predicted by the lightcurve/lifetime models in 
    \citet[][lines]{hopkins:qso.all,hopkins:faint.slope,hopkins:bol.qlf}. Solid red line is the simplest 
    model prediction for a population of 
    single triggers with $t_{\rm Q}\ll t_{H}$; black lines integrate over a complete 
    cosmological history of triggering events in the model (constrained to match the observed 
    AGN luminosity functions). Dashed and dotted lines bracket the model uncertainty. 
    The lines are {\em predictions} -- there are {\em no} free parameters fitted or tuned to 
    match the observed $\lambda$ distributions. 
    \label{fig:yulu.1}}
\end{figure*}

\begin{figure*}
    \centering
    \plotone{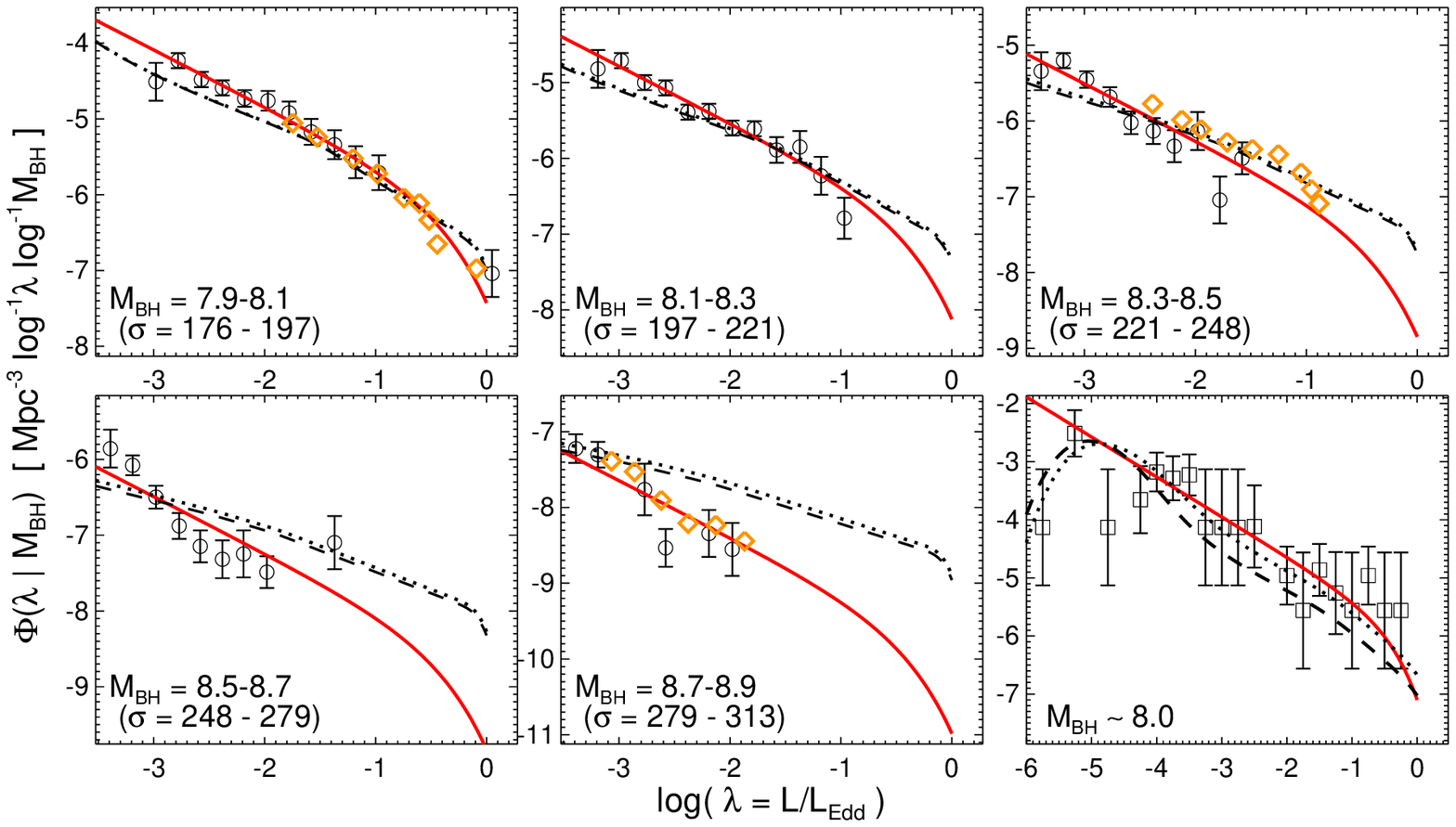}
    \caption{As Figure~\ref{fig:yulu.1}, continued to higher BH masses. 
    The lower right panel shows the Eddington ratio distribution calculated in 
    \citet{hopkins:old.age} from the observed samples of \citet{marchesini:low.mdot.sample} and 
    \citet{ho:radio.vs.mdot}, 
    with a less tightly constrained BH mass range ($M_{\rm BH}\sim10^{7.5}-10^{8.5}\,\msun$) 
    but extending to lower Eddington ratios. Corrections for various changes in radiative 
    efficiency with Eddington ratio change the observed points and models within the 
    plotted error bars \citep[see][]{hopkins:old.age}. Note that the apparent discrepancy in 
    the models at the highest masses comes from extrapolation to unobservably small 
    space densities. 
    \label{fig:yulu.2}}
\end{figure*}

Figures~\ref{fig:yulu.1}-\ref{fig:yulu.2} show this analysis. We plot the Eddington 
ratio distribution determined in \citet{yu:mdot.dist} 
and \citet{heckman:local.mbh} \citep[see also][]{kauffmann:new.mdot.dist}, 
and compare to the theory of 
\citet{hopkins:qso.all,
hopkins:faint.slope,hopkins:bol.qlf}. We consider three versions of the 
estimates to show the 
range of theoretical uncertainty inherent in the model, but the differences are, 
for the most part, minimal. First, we plot the \citet{hopkins:faint.slope} fitted $\dtdlogl$ distribution 
to their typical $\sim M_{\ast}$ BH simulations, multiplied by the number density of 
active BHs at each mass (the effective duty cycle, 
discussed in \S~\ref{sec:compare.models}) 
and divided by the Hubble time (since we are considering the 
fraction of active BHs at each $\mdot$; this is shown as a the solid red line). This 
is an accurate approximation to a much more complete cosmological calculation 
so long as the quasar lifetime is short compared to the Hubble time and 
provided that 
there is not some strong feature in the redshift history of triggering (i.e.\ so long 
as the cosmological evolution around the redshift of interest is relatively weak). 
At low $\mdot$, this is simply a power-law; inevitably, the point where 
the lifetime nears the Hubble time will be reached, and more ful cosmological 
models will turn over.

Second, we adopt the fits from \citet{hopkins:bol.qlf} to the observed bolometric 
QLF as a function of redshift, using the lifetime distributions as a function of BH 
mass fitted in \citet{hopkins:faint.slope}, and integrate these over redshift. 
In \citet{hopkins:bol.qlf} the authors quantify the range of fits allowed given both the 
uncertainty in the lifetime model and the observed QLF; we show two lines that 
bracket this range (the dotted and dashed lines in Figures~\ref{fig:yulu.1}-\ref{fig:yulu.2}). 
Again, these use the same model lightcurves \citep[the power-law like fits to the 
lightcurve shape as a function of BH mass, from][]{hopkins:faint.slope} 
and are matched to the same QLF \citep[the data compiled in][]{hopkins:bol.qlf}. 
However, there remain uncertainties in the data and degeneracies in the fit -- the 
models shown bracket the range in the 
$\mdot$ distribution allowed by this set of models and data. 
The predicted distributions turn over at low $\mdot$ in both cases (albeit at 
slightly different $\mdot$, reflecting these degeneracies in fitting the low-luminosity 
population), as the lifetime nears the Hubble time (obviously, in a full cosmological 
model, they must turn over, so that the ``duty cycle'' of a given BH across all 
Eddington ratios integrates to unity). 

The theoretical predictions all agree well, and agree remarkably well with the observations. 
We stress that these are {\em predictions}; no quantity has been fitted to the data points 
(both the shape and normalization of the model curves are entirely predicted by the 
papers above). 

The theoretical curves also agree with one another. This is because, at most of the 
luminosities of interest, the lifetime is well below the Hubble time and 
the (especially low-mass) BHs are still growing as a population 
\citep[see e.g.][]{hasinger05:qlf}, so subtleties of 
cosmological evolution are irrelevant (the only quantity of interest is $\dtdlogl$). 
Moreover at low redshifts $z\lesssim2$ the observed QLF is well-constrained, so there 
is little freedom given a $\dtdlogl$ model in what the distribution can be. 
The only differences appear at the highest masses -- this is because, in a complete 
cosmological calculation, most of the triggering of these systems occurred preferentially at high 
redshifts, so there is some pile-up at low Eddington ratios (precisely below where the 
condition of $\tq\sim t_{H}$ begins to be satisfied, as expected). 

Figure~\ref{fig:yulu.2} also compares the Eddington ratio distribution determined 
in \citet{hopkins:old.age} from the combined observed samples of \citet{marchesini:low.mdot.sample} 
and \citet{ho:radio.vs.mdot}. The range in BH mass in this sample is less narrowly constrained, 
$M_{\rm BH}\sim 10^{7.5}-10^{8.5}\,\msun$, but the data survey extremely faint systems in 
the radio and X-rays, and allow us to extend the observed Eddington ratio distribution from 
the already deep $\mdot\sim10^{-4}-10^{-3}$ in the \citet{yu:mdot.dist} and 
\citet{heckman:local.mbh} samples 
by another two orders of magnitude. (Note in \citet{hopkins:old.age} the authors consider 
possible corrections for varying radiative efficiencies and bolometric corrections 
at these luminosities, but the resulting change is for our purposes here within the error bars 
shown.)  Again, the agreement is good over the entire range.

\begin{figure*}
    \centering
    \plotone{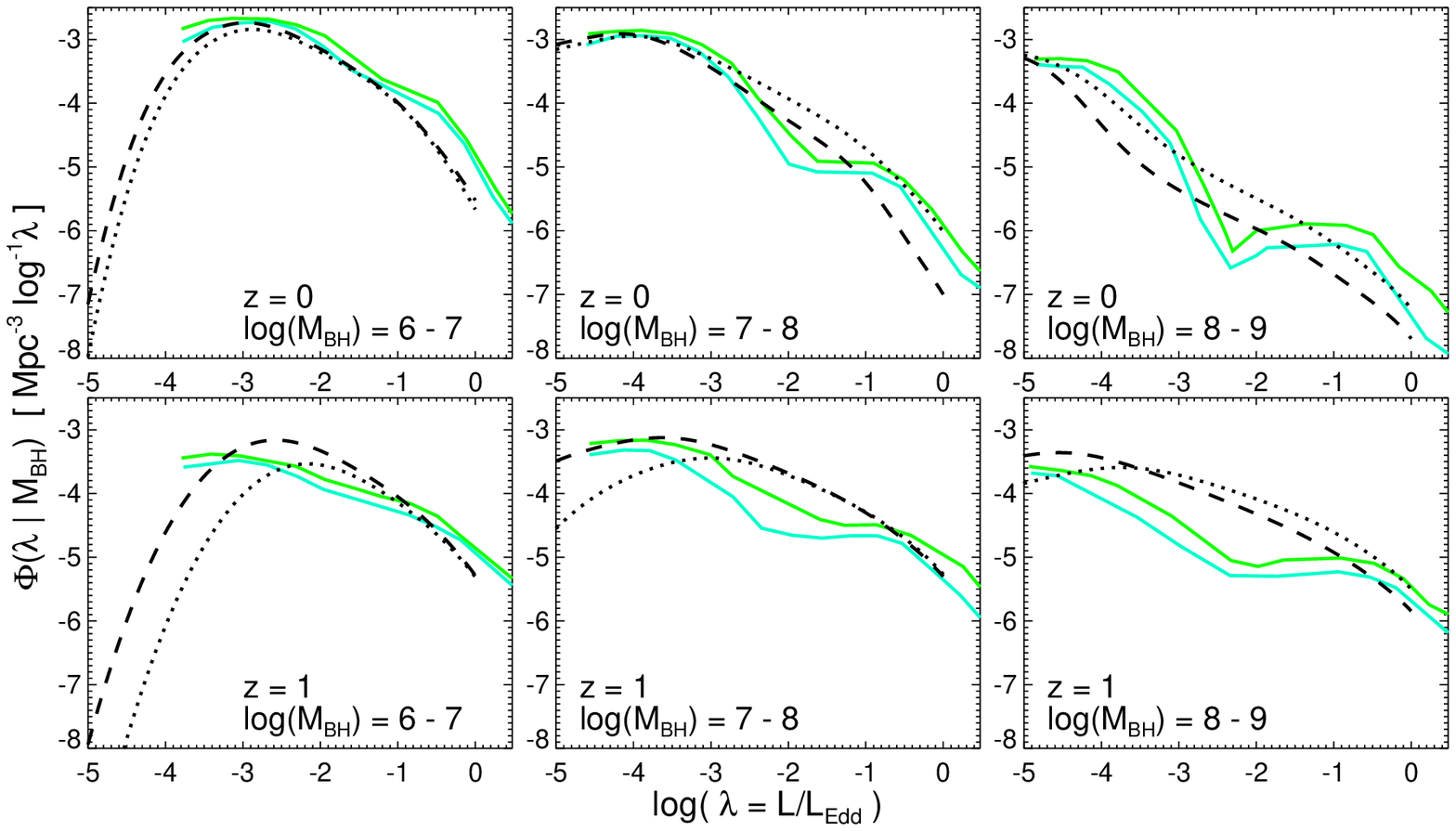}
    \caption{As Figure~\ref{fig:yulu.1}, but comparing the model predictions with 
    the observationally inferred Eddington ratio distribution from the combination of 
    observed X-ray, radio, and optical luminosities in \citet{merloni:synthesis.model,
    merloni:mdot.dist.fits.prep} and the 
    statistical correlation between e.g.\ radio-loudness and accretion 
    rates \citep[for a review see][]{fender:radio.mdot.review}.
    The results are shown over the plotted mass bins, at 
    each labeled redshift. The sharp kinks in the \citet{merloni:mdot.dist.fits.prep} distributions are 
    somewhat model-dependent. 
    \label{fig:merloni}}
\end{figure*}

At higher redshifts, analogous measurements are not, at present, available. However, 
we can compare with an alternative, albeit indirect, observational estimator. 
It has been argued that the combination of radio, X-ray, and optical luminosities 
can be used to constrain the Eddington ratios of BHs, in a manner like that
well-established for accreting X-ray binaries.
Indeed, it is increasingly 
established that radio-loudness of AGN appears to be a function (on average) of 
Eddington ratio \citep[][for a review see \citet{fender:radio.mdot.review}]{nym95,nym96,
falcke:radio.vs.mdot,meier:jets.in.adaf,
ho:radio.vs.mdot,merloni:bhfp.radio,marchesini:low.mdot.sample,
maccarone:agn.riaf.connection,falcke04:radio.vs.mdot,
merloniheinz.bhfp.radio,greene:radio.vs.edd}, so if this is true at higher redshift it can be used as at 
least a statistical estimator of Eddington ratio. \citet{merloni:synthesis.model} adopt these observations,
combined with the measured X-ray, radio, and optical luminosity distributions of 
observed quasars, to constrain the Eddington ratio distribution as a bivariate function of 
BH mass and redshift. Their methodology allows for intrinsic scatter in these correlations, so 
it should be reasonably robust as long as there is some physical relationship between 
Eddington ratio and X-ray-radio-optical spectral shape over a wide baseline in Eddington ratio 
(from $\sim10^{-5}-1$). 

Figure~\ref{fig:merloni} compares their inferred Eddington ratio distributions 
\citep[A.\ Merloni, private communication; for details see][]{merloni:mdot.dist.fits.prep}
to the same predictions, at $z=0$ and $z=1$ (at higher redshifts, 
\citet{merloni:synthesis.model,merloni:mdot.dist.fits.prep} 
do infer Eddington ratio distributions, but the radio luminosity function 
is not directly measured and the X-ray luminosity function is increasingly unconstrained as 
well, so this relies on extrapolation of the low-redshift trends, and 
is not a direct measurement/estimate). Note that the kink in the \citet{merloni:synthesis.model} 
Eddington ratio distributions around $\mdot\sim10^{-2}$ is 
sensitive to the particular assumptions about the form of the QLF 
shape and SED shape as a 
function of luminosity; other attempts to infer this distribution have, however, 
seen similar features \citep[e.g.][]{marchesini:low.mdot.sample}, so it may reflect a 
real change in accretion properties not modeled here. In any case, the 
inferred distributions, where they overlap with the measurements from 
the observations above, agree reasonably well, giving some confidence in this 
methodology, and the agreement with theoretical predictions is good.

As discussed in \S~\ref{sec:intro}, although the particular models shown 
were fitted from simulations of galaxy mergers, the results are comparable regardless of 
fueling mechanism, given a similar self-regulation from local AGN feedback. 
In fact, the predictions and their agreement with 
observations in Figures~\ref{fig:yulu.1}-\ref{fig:merloni} 
rely on {\em no} information regarding fueling mechanisms -- 
they simply follow from assuming a given lightcurve shape 
and matching the observed quasar luminosity function.

\section{Eddington Ratio Distributions as a Function of Luminosity} 
\label{sec:compare.L}

At a given luminosity (as opposed to a given BH mass), the 
differences in the Eddington ratio distribution between various models 
are greatly suppressed. The reasons for this are obvious: at a given 
$L$ and $M_{\rm BH}$, there is only a narrow range of $\mdot$, and given the 
declining number density of high-mass BHs, at high-$L$, one will increasingly 
be limited to the population of near-Eddington systems. However, over a sufficiently 
large baseline in $L$, differences are apparent, and (especially at high 
redshifts) luminosity-limited samples are more easily constructed 
than $M_{\rm BH}$-limited samples. We therefore consider the Eddington ratio 
distribution predicted as a function of luminosity at different redshifts. 

\begin{figure}
    \centering
    \plotone{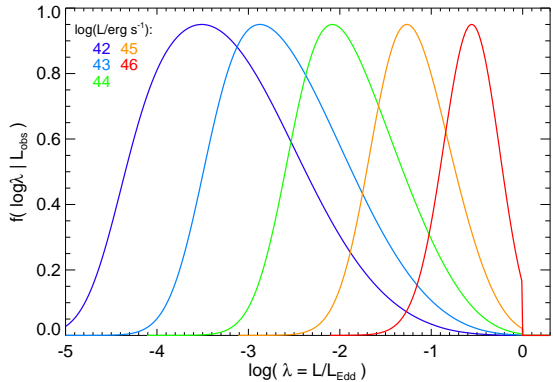}
    \caption{Predicted distribution of Eddington ratios at $z\sim1$ in a narrow 
    range of bolometric luminosity. Note that, because of the cut at a given luminosity 
    (as opposed to reflecting all $\lambda$ at a given $M_{\rm BH}$ as in 
    Figure~\ref{fig:yulu.1}), 
    the distributions are much more narrow and turn over at low $L_{\rm bol}$. 
    The distributions are reasonably approximated as log-normal, with some 
    weakly luminosity-dependent skewness. 
    \label{fig:mdot.at.L}}
\end{figure}

\begin{figure*}
    \centering
    \plotone{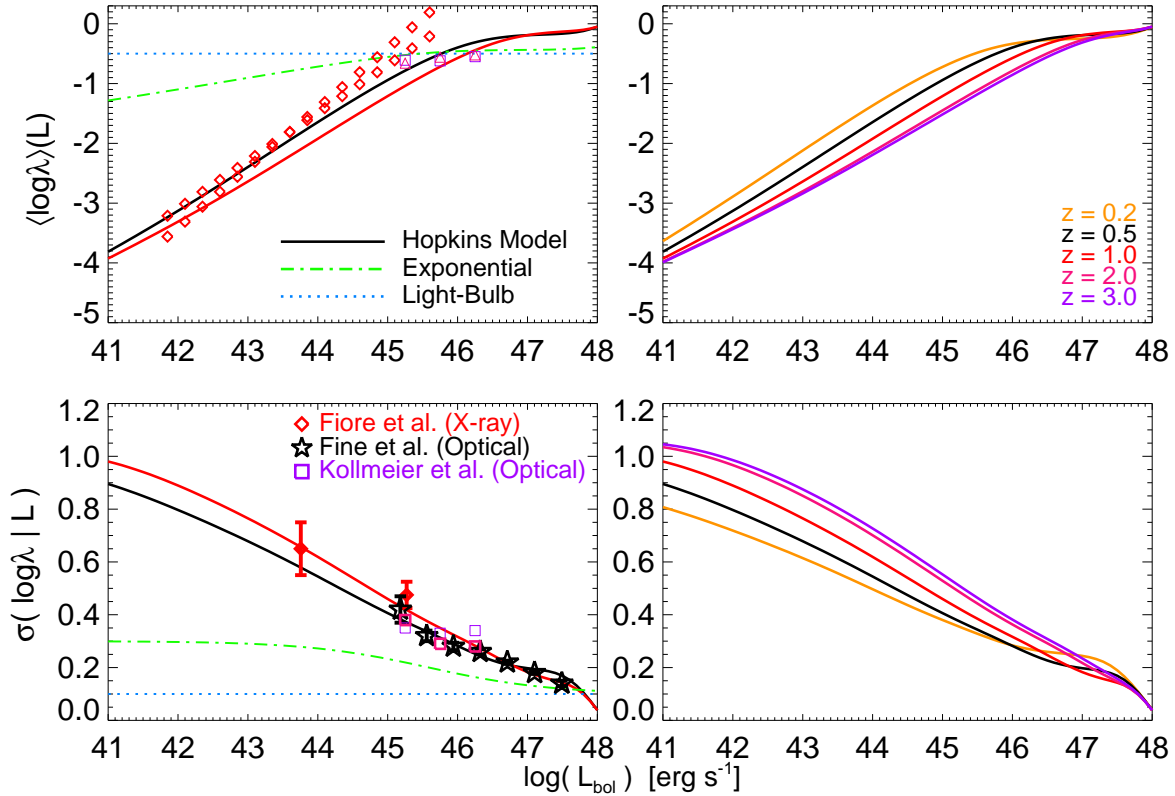}
    \caption{Median Eddington ratio $\lambda$ and $1\,\sigma$ dispersion 
    in Eddington ratios at a given narrow range in bolometric AGN luminosity 
    (as in Figure~\ref{fig:mdot.at.L}, assuming a lognormal distribution). 
    We compare the \citet{hopkins:qso.all,hopkins:faint.slope,hopkins:bol.qlf} 
    model predictions (solid lines; colors denote redshift as labeled)
    with the observed distributions inferred from the distribution of 
    X-ray to host luminosities in optically obscured AGN 
    \citep[][red diamonds]{fiore:type2.lx.vs.lhost,hasinger:absorption.update,
    hickox:bootes.obscured.agn,hickox:multiwavelength.agn} (translating host luminosity to 
    average BH mass given the observed correlations), and 
    from the distribution fitted to broad-line optical samples 
    in \citet{fine:broadline.distrib} from the 2dF and \citet{kollmeier:mdot} from AGES, using 
    the optical virial (line-width) BH mass estimators. 
    We also compare with the predictions for pure exponential 
    AGN lightcurves (constant Eddington ratio or exponential decay in $L$ 
    after some peak) and ``light-bulb'' models (where AGN are ``on'' or 
    ``off'' with a mass-independent narrow $\lambda$ distribution when on), 
    forced to obey the (necessary) constraint of matching the observed AGN 
    luminosity functions. 
    The \citet{hopkins:faint.slope} models agree well with the observations, the other 
    models are ruled out at low-L. High-L optical samples are not ideal for 
    breaking the degeneracies between these models, although with 
    sufficiently large dynamic range such as that in \citet{fine:broadline.distrib} 
    the distinction can be seen. 
    \label{fig:mdot.by.L}}
\end{figure*}

Figures~\ref{fig:mdot.at.L} \&\ \ref{fig:mdot.by.L} shows the results. 
Figure~\ref{fig:mdot.at.L} demonstrates that at a given luminosity (unlike at a 
given BH mass), the Eddington ratio distribution is expected to be something like a 
lognormal distribution (more so if typical observational errors are included), albeit 
with some non-negligible skewness \citep[as seen 
in e.g.][]{fine:broadline.distrib}.\footnote{The skewness 
originates because the distribution of host BHs is not flat in mass, but increases 
to lower masses (following the Schechter galaxy/spheroid mass function). With respect 
to the median Eddington ratio/BH mass contributing to the observed 
population at a given luminosity, one therefore expects (if Eddington ratio distributions at a given 
BH mass do not change rapidly with mass) that there will be a somewhat larger population 
of low-mass BHs at high Eddington ratio rather than high-mass BHs (an exponentially 
vanishing population) at low Eddington ratio.} 
This is because, 
at extremely low $\mdot$, arbitrarily high-$M_{\rm BH}$ BHs would be implied (in a bin 
of fixed $L$), but the possible population of such systems is vanishing. 
We therefore find it convenient to approximate the predicted distributions as lognormal, 
and quantify the median $\mdot$ and $1\,\sigma$ dispersion (technically based 
on the IPV width to prevent bias from outliers or skewness) 
in the lognormal as a function of luminosity, at redshifts $z=0-2$. 

We compare these with several observational estimates. In \citet{fine:broadline.distrib}, the 
authors consider the Type 1 quasar population near $z\approx1$ 
in the 2dF survey, and estimate the distribution of BH masses in narrow bins 
of luminosity employing the commonly adopted 
virial BH mass estimators \citep[based on the broad-line 
widths and the radius-luminosity relations inferred from reverberation mapping 
of nearby AGN; see e.g.][and references therein]{vestergaardpeterson:virial.corr.review}. 
This allows them to consider the distribution of 
BH masses via this proxy as a function of luminosity down to near Seyfert luminosities. 
Because the width of the distribution can be determined without relying on the 
(systematically still uncertain) absolute normalizations of these calibrators, the 
authors decline to estimate absolute Eddington ratios (although a rough estimate 
suggests they lie between $\sim0.1-1$, as predicted here). 
\citet{kollmeier:mdot} use the same technique over a narrow luminosity range 
of Type 1 AGN in the AGES survey. 

At lower luminosities, these indicators 
are less useful (and the observations suggest the population is both more 
obscured and diluted by host galaxy light, making the virial mass estimators 
inaccessible). However, X-ray observations can probe Type 2 objects in this 
regime, where the optical light is dominated by the host galaxy and
therefore a host galaxy stellar mass \citep[and corresponding BH mass, 
adopting the observed $M_{\rm BH}-M_{\ast}$ relation from][]{marconihunt} 
can be estimated.
It is well-established that in this regime the optical luminosity/stellar 
mass of the galaxy is approximately constant while the X-ray luminosity changes, 
implying that at lower-$L$ the X-ray luminosity function becomes increasingly 
a sequence in Eddington ratio. 

We plot the implied Eddington ratios and 
distribution in Eddington ratios as a function of luminosity from the 
sample of \citet{hasinger:absorption.update}, where the optical $R$-band luminosity is converted to 
a stellar mass based on the age and mass-dependent observed mean $M/L$ 
ratios in \citet{belldejong:disk.sfh} and \citet{bell:mfs}. We have re-calculated these 
comparisons using e.g.\ the samples of \citet{fiore:type2.lx.vs.lhost} and 
\citet{hickox:bootes.obscured.agn,hickox:multiwavelength.agn} and 
obtain the same result (various multiwavelength surveys have 
reached similar conclusions regarding this correlation), 
and find that changing the assumed host $M/L$ within 
uncertainties makes little difference \citep[for more 
discussion, see][]{hopkins:seyfert.bimodality}. For convenience to
the comparison here, we convert all 
the observed AGN luminosities to bolometric luminosities using the bolometric 
corrections in \citet{hopkins:bol.qlf} \citep[using instead those in][makes no 
difference]{elvis:atlas,marconi:bhmf,richards:seds}.

In addition to the theoretical predictions from the models in \S~\ref{sec:compare.dist}, 
we contrast the results from a simple light-bulb AGN lifetime
(in which both the mean Eddington ratio and dispersion are constant) 
and a pure exponential model (in which $\dtdlogl$ is constant at $\mdot\ll1$, 
which, when the sample is cut by luminosity, does introduce some dependence 
on $L$, but much weaker than that predicted by the models in \S~\ref{sec:compare.dist}). 
The observations clearly prefer the steeper dependence of the 
more realistic lifetime models. Note that a large baseline in luminosity is 
needed to see the difference at high significance -- at least $\sim2-3$ orders of 
magnitude in $L$ below $L_{\ast}$ (ideally more like $\sim4-5$ orders of magnitude). 
This explains e.g.\ the weak dependence seen in the sample of \citet{kollmeier:mdot}, 
who note the weak dependence of mean Eddington ratio and width of the distribution on 
$L$, but have a relatively narrow range in $L$ and are concentrated near 
and above $\sim L_{\ast}$. Their observations are in fact entirely consistent with the 
model predictions and other observations (over larger baselines) that do see 
such a dependence.

\section{Comparison with Models: What Degeneracies Are Broken?} 
\label{sec:compare.models}

We now ask how unique these predictions are: in other words, can the observations 
distinguish between different models for the Eddington ratio/quasar lifetime distribution?

\begin{figure*}
    \centering
    \plotone{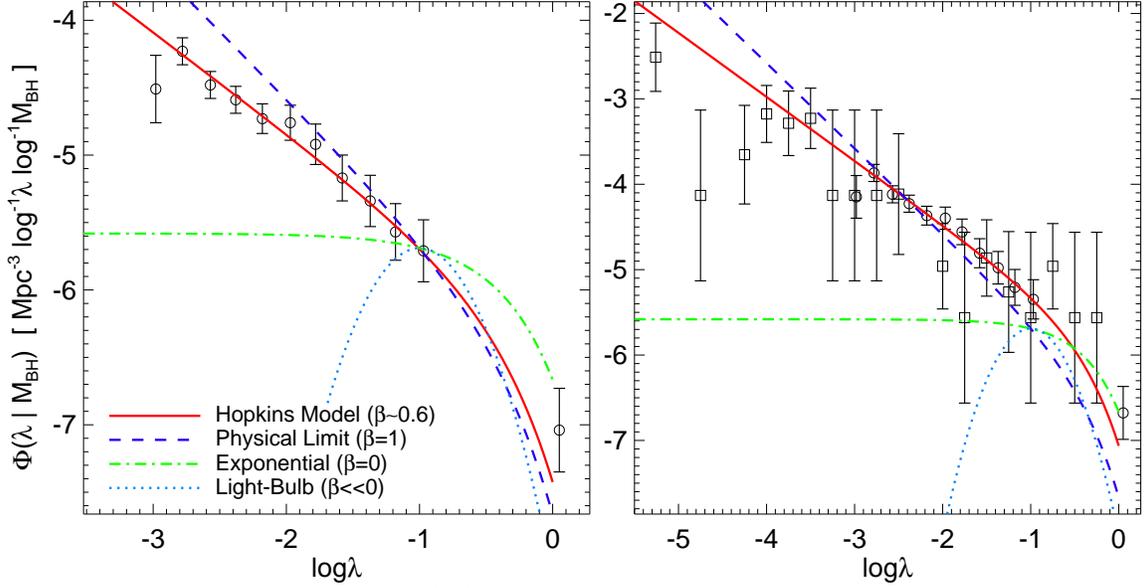}
    \caption{As Figure~\ref{fig:yulu.1}, but comparing the consequences of 
    different lightcurve/lifetime models for the observed Eddington ratio distribution 
    (generally parameterized by Equation~\ref{eqn:schechterfit}, with power-law like slope 
    $-\beta$ at low $\lambda$). We compare our previous model examples 
    (red solid, with median $\beta\sim0.6$) and the exponential 
    (green dot-dashed, $\beta=0$) and light-bulb (blue dotted, $\beta\ll0$, reflecting a 
    log-normal or $\delta$-function distribution) models. We also show the effective 
    physical upper limit (dark blue dashed, $\beta=1$; $\beta>1$ would result in BH 
    growth that is divergent towards low-$\lambda$). The models are all normalized 
    to give the same number of objects at $\lambda\gtrsim0.1$; this is required
    in order to match the observed quasar luminosity function where it is well-constrained 
    ($L \gtrsim 10^{45}\,{\rm erg\,s^{-1}}$). Even freeing this normalization (i.e.\ allowing the 
    AGN luminosity function to be significantly different from that observed), however, still 
    does not allow the exponential or light-bulb models to match the observations. 
    \label{fig:model.compare}}
\end{figure*}

Figure~\ref{fig:model.compare} compares the data and several simple models for the 
lifetime distribution. There are a number of commonly adopted forms
for quasar lifetimes and lightcurves in the 
literature, including the light-bulb, 
pure exponential growth (corresponding to growth at fixed Eddington ratio and 
then either instantaneous decline or similar decay), and self-regulated 
decay models. The range of 
possibilities can be generally approximated 
by a Schechter fitting function: 
\begin{equation}
\frac{{\rm d}t}{{\rm d}\log{L}} = t_{0}\,
{\Bigl(}\frac{L}{L_{\rm peak}}{\Bigr)}^{-\beta}\,
\exp{(-L/L_{\rm peak})}
\label{eqn:schechterfit.0}
\end{equation}
where $L_{\rm peak}\equiv\eta\,L_{\rm Edd}$ with some $\eta\sim1$. 
Equivalently, we can fit: 
\begin{equation}
\frac{{\rm d}t}{{\rm d}\log{\mdot}}(M_{\rm BH}) = t_{0}\,
{\Bigl(}\frac{\mdot}{\eta}{\Bigr)}^{-\beta}\,
\exp{(-\mdot/\eta)}.
\label{eqn:schechterfit}
\end{equation}

This allows for the fact that there must be some physical (and observed) 
cutoff in the Eddington ratio distribution above $\mdot=1$, but permits
an arbitrary power-law like behavior at lower Eddington ratios 
(which can approximate any near-similarity solution for self-regulated 
lightcurve decay, or a power-law-like spectrum of triggering 
activity, as well as a rapid/exponential rise fall and even sufficiently rapid 
rise similar to a light-bulb). In practice, we find that fitting 
an equation of the form of Equation~\ref{eqn:schechterfit} to the 
data in e.g.\ Figures~\ref{fig:yulu.1}-\ref{fig:merloni} provides a 
useful and statistically good description of the data. 

In all mass intervals, we 
find a similar best-fit $\eta\approx0.2-0.4$ -- the exact value 
reflects the particular choice of functional 
form, but the general value describes the 
observed and expected cutoff in the $\mdot$ distribution at 
$\mdot\sim1$. We discuss variations in $\eta$ below, but find that 
it cannot vary widely: much lower $\eta$ exponentially suppresses 
the number of bright sources, 
and much higher $\eta$ implies no Eddington limit, both 
in conflict with AGN luminosity functions and direct observations of 
the distribution of accretion rates in bright, broad-line systems 
\citep[see e.g.][]{kollmeier:mdot,greene:active.mf}. In what follows, 
we find statistically identical results fitting the observations with 
a free $\eta$ or fixed $\eta=0.4$ (the degeneracy between e.g.\ $\eta$ 
and $\beta$, within the range allowed by observations, is not strong). 

As a consequence, 
the shape of the distribution is primarily contained in the slope $\beta$. 
Various lightcurve models make differing specific predictions for 
this slope: 
\\

{$\bullet$ \bf Light-Bulb Models:} Strictly speaking, 
in such a model $\dtdlogl$ is a delta function at the characteristic 
$\mdot_{\rm on}$; allowing for some finite width or measurement errors in 
$\mdot$, this can be approximated as e.g.\ a lognormal distribution or a 
Schechter distribution with $\beta \ll 0$ (i.e.\ a large {\em negative} $\beta$). 

{$\bullet$ \bf Exponential Models:} For pure exponential models (e.g.\ accretion at a constant 
Eddington ratio or exponential decay), 
one obtains $\dtdlogl\approx$constant, or $\beta=0$. 

{$\bullet$ \bf Maximal low-$\lambda$ Accretion:} 
At the opposite extreme, there is a physical limit $\beta < 1$; because the fractional contribution 
to BH mass growth from a range in $\log(\mdot)$ 
goes roughly as $\mdot\times\dtdlogl$.  For $\beta\ge1$, the total growth is 
both formally divergent and (even if there is some cutoff at low-$\mdot$)
weighted towards the lowest-$\mdot$ values. 
Constraints from the observed BH mass function and the 
\citet{soltan82} argument \citep[see e.g.][]{yutremaine:bhmf,
salucci:bhmf,shankar:bhmf,marconi:bhmf}, as well as other 
indirect constraints \citep{hopkins:old.age} imply that most BH growth cannot 
occur in extremely low Eddington ratio states. 

{$\bullet$ \bf Self-Regulated Models:} In feedback-regulated models, 
the energy coupled to the ISM which halts quasar accretion leads 
to a nearly self-similar power-law like decay of the quasar lightcurve, 
$L\propto t^{-1/\beta}$. Hydrodynamic simulations of 
quasars in galaxy mergers \citep{hopkins:faint.slope}  
suggest values $\beta\sim0.6$ for typical $\sim L_{\ast}$ 
galaxies, with a weak mass dependence. Analytic calculations 
in \citet{hopkins:seyferts} demonstrate that for a range of assumptions 
regarding the fueling 
mechanisms (for e.g.\ stochastic or secular AGN fueling 
mechanisms), feedback coupling mechanism, timescale, efficiency, and 
gas properties around the BH, a range of $\beta\approx0.3-0.8$ is possible; 
but simulations of these non-merger scenarios \citep{younger:minor.mergers} 
suggest a similar, relatively narrow range of $\beta$ independent of 
fueling mechanism. 

{$\bullet$ \bf Isolated Accretion Disk/Gas Starvation Models:} \citet{yu:mdot.dist,yulu:lightcurve.constraints.from.bhmf.integration} show 
that a similar power-law decay is expected 
for a thin $\alpha$-disk \citep{shakurasunyaev73} abruptly 
cut off from any future fuel supply, when there is no feedback. 
The disk therefore slowly starves as gas exhausts by 
a combination of accretion and (possibly) star formation. 
The solution should be mass and host galaxy independent, with $\beta\approx0.80-0.84$. 
The exact solution depends (weakly) in detail on e.g.\ how 
the opacity and viscosity of the disk vary as a function of density and temperature, but the 
authors show that the entire set of solutions (corresponding to the allowed 
range for observationally reasonable $\alpha$-disks) lie within this narrow 
interval in $\beta$. In Figure~\ref{fig:model.compare}, we omit this 
model for clarity, but the result lies between the self-regulated 
model and the physically maximal ($\beta=1$) model. 

{$\bullet$ \bf BHs Trace Stellar Evolution:} It is possible that many low-luminosity 
systems are fueled by stellar mass loss from aging nuclear stellar populations. 
If the BH simply linearly traced this evolution (i.e.\ grew in this fueling-limited 
background with no feedback), the late-time accretion rate evolution would 
follow the stellar mass loss rate, giving a similar power-law 
solution with a steeper $\beta\approx0.9-1.0$ \citep[e.g.][]{norman:stellar.wind.fueling,
starburst99,BC03,ciottiostriker:recycling}. \\

\begin{figure}
    \centering
    \scaleup
    \plotter{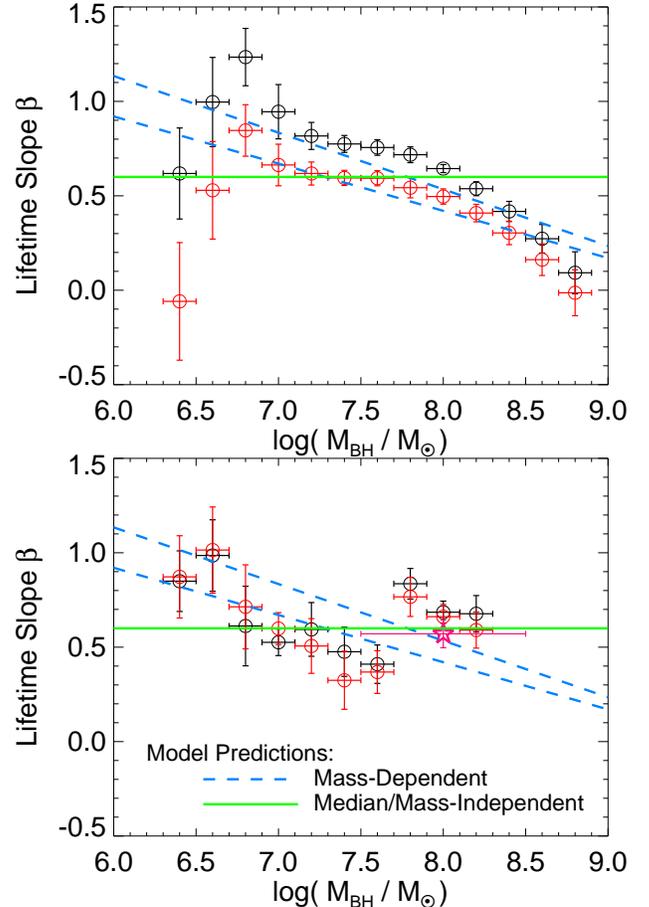}
    \caption{Slope $\beta$ of the lifetime/lightcurve/Eddington ratio distribution 
    fitted to the observations in Figures~\ref{fig:yulu.1}-\ref{fig:yulu.2} with a 
    general Schechter-function parameterization (Equation~\ref{eqn:schechterfit}). 
    {\em Top:} Points are the maximum likelihood 
    fit results for each mass bin, where the fits are 
    constrained such that the observed quasar luminosity function must 
    be reproduced. We repeat our fits using the quoted error bars 
    in the observed Eddington ratio distributions (black) and 
    allowing for an additional $\sim0.3$\,dex intrinsic uncertainty (red). 
    We compare with the model predictions from \citet{hopkins:faint.slope}, 
    determined from simulations 
    allowing for a mass-dependent typical lifetime distribution or 
    assuming a constant (mass-independent) lifetime distribution ($\beta\approx0.6$). 
    {\em Bottom:} Fits repeated, but removing the constraint that the observed 
    luminosity function must be reproduced (magenta star is the fit to the 
    observations in the lower-right panel of Figure~\ref{fig:yulu.2}, which span a 
    wider mass and $\mdot$ interval). The results are consistent, 
    but less constrained. 
    The results agree well with the model predictions, and if a match to the 
    observed luminosity function is required, specifically favor the 
    full mass-dependent model. In either case, an exponential ($\beta=0$) or 
    light-bulb ($\beta\ll0$) model is ruled out at high significance.  
    \label{fig:pwrlaw}}
\end{figure}

Fitting each BH mass bin in Figures~\ref{fig:yulu.1}-\ref{fig:yulu.2} to 
an arbitrary 
function of the form in Equation~\ref{eqn:schechterfit} (free $t_{0}$, $\eta$, $\beta$), we 
obtain the constraints on $\beta$ shown in Figure~\ref{fig:pwrlaw}. 
In fact, the constraints can be made stronger. Any fitted $\lambda$ distribution, 
convolved with the BH mass function \citep[implicit in 
Figures~\ref{fig:yulu.1}-\ref{fig:yulu.2}, or taken from observations 
following e.g.][]{marconi:bhmf} must reproduce the observed AGN luminosity 
functions. In practice, this ``anchors'' the number density at high $\lambda\gtrsim0.1$ 
(which dominate the QLF). Re-fitting the observed $\lambda$ distribution, with the 
fit constrained to also reproduce the observed QLF 
\citep[taken here from the compilation of a large number of observations in][but 
the results are not sensitive to the specific choice]{hopkins:bol.qlf}, we obtain the 
constraints in the top panel of Figure~\ref{fig:pwrlaw} (note that the errors  
shown in this case are not independent). 

Parametrically, if we assume $\beta$ is independent of $M_{\rm BH}$ 
(or consider a range near $\sim L_{\ast}$), we obtain 
$\beta \approx 0.6\pm 0.05$, in good agreement with the theoretical predictions 
from the simulations of \citet{hopkins:faint.slope}. 
If we allow for a dependence of $\beta$ on $M_{\rm BH}$, parameterized 
for convenience as 
\begin{equation}
\beta = \beta_{7.5} + \beta^{\prime}\,\log{(M_{\rm BH}/10^{7.5}\,\msun)}, 
\label{eqn:beta.vs.mbh}
\end{equation}
we obtain $\beta_{7.5}=0.63\pm0.04$ and $\beta^{\prime}=-0.29\pm0.08$ if we 
include the constraints from the observed QLF, and 
$\beta_{7.5}=0.63\pm0.04$ and $\beta^{\prime}=-0.11\pm0.09$ if 
we do not (the observed $\lambda$ distributions themselves are insufficient 
to determine if $\beta$ depends weakly on BH mass).

\begin{figure}
    \centering
    \scaleup
    \plotone{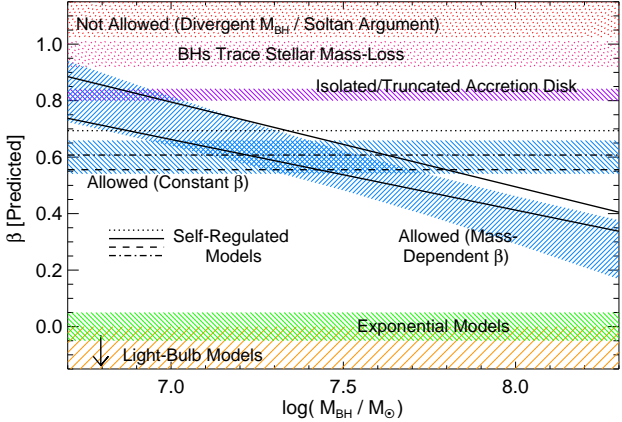}
    \caption{Illustration of how the observational constraints on $\beta$ can 
    discriminate between different physical and phenomenological quasar 
    lifetime models. Blue shaded range shows the allowed slope $\beta$ from 
    the fits to the observations taken cumulatively, either assuming it 
    is invariant (independent of $M_{\rm BH}$) or allowing it to depend on 
    $M_{\rm BH}$ as in Figure~\ref{fig:pwrlaw}. Black lines compare different 
    models for the self-regulating growth seen in hydrodynamic simulations 
    \citep[where e.g.\ the gas density profiles, ISM equation of state and phase 
    breakdown, and prescriptions for feedback and BH accretion are varied; for details 
    see][]{hopkins:faint.slope,hopkins:seyferts}; these predictions agree well with the observations 
    and are relatively independent of the detailed assumptions. If we include 
    the constraint that the QLF must be reproduced simultaneously, the observations 
    rule out mass-independent $\beta$ (corresponding to slightly 
    more simplified analytic self-regulated 
    models, described in the text). 
    The $\beta$ corresponding to pure exponential or phenomenological ``light-bulb'' 
    models are also shown, as is the region ruled out by the \citet{soltan82} and other 
    physical arguments. An alternative physical model, of a thin $\alpha$-disk where 
    the entire fuel supply is instantaneously provided from larger radii then cut off 
    (i.e.\ large-scale gas inflows and feedback are excluded), is shown 
    \citep[this yields $\beta\approx0.8$ as the 
    accretion disk consumes gas; see][]{yu:mdot.dist}. Although a better description than the simplified 
    phenomenological models, the observations can rule out such strictly ``isolated'' 
    accretion disk exhaustion at $\gtrsim3\,\sigma$ (the constraints are even stronger, 
    $\sim5\,\sigma$ at high masses, if we also force the solution to match the observed QLF). 
    Likewise, assuming BH accretion simply traces stellar mass loss from evolving 
    stellar populations (without feedback to regulate growth more efficiently) 
    is ruled out ($\beta\approx0.9-1.0$). 
    \label{fig:beta.models}}
\end{figure}

Figure~\ref{fig:beta.models} summarizes the predictions of the 
different models above. 
The observations prefer a narrow range of slopes: it is 
not possible to match the observed Eddington ratio 
distribution in light-bulb or exponential models (ruled out at $\gg 5\,\sigma$), 
and simple stellar wind models are ruled out at $\sim4\,\sigma$. 

The isolated accretion disk model fares somewhat better: 
\citet{yu:mdot.dist} show that it is consistent with the $\lambda$ distribution observed 
for low-mass BHs, a result we confirm here, but it is inconsistent with 
the observations of typical $\sim L_{\ast}$ and more massive 
BHs at $>4\,\sigma$. 
\citet{yulu:lightcurve.constraints.from.bhmf.integration} 
show that such a model 
does improve upon the light-bulb model in simultaneously reproducing 
the observed QLF and BH mass function. However, as noted before, the QLF is 
primarily a tracer of high-$\lambda$ activity, and given that such activity dominates 
the accretion history of BHs \citep{soltan82}, these observational constraints 
alone cannot break the degeneracy between similar, but slightly different 
lightcurve models $L\propto t^{-1/\beta}$ ($\beta\sim0.6$ and $\beta\sim0.8$ 
corresponding to the self-regulated and isolated accretion disk models, respectively). 
\citet{hopkins:qso.all,hopkins:bol.qlf,hopkins:groups.qso,hopkins:groups.ell} 
demonstrate that similar solutions for the QLF and BH mass function exist for 
the self-regulated model (unsurprisingly, the two yield very similar growth histories 
and $\lambda\gtrsim0.1$ activity). As a consequence 
the low-$\lambda$ distribution represents an important 
means to break these degeneracies, and disfavor 
the isolated accretion disk models. We do emphasize, however, that 
while the isolated disk model is
ruled out at high formal significance, the absolute 
difference between it and self-regulated models
is not large. That modest difference may reflect the inclusion of a small, 
but physically important term, for example outflows or winds 
from the accretion disk that cause it to gas-exhaust slightly more 
steeply in time than otherwise. 

Feedback-regulated models agree well with observations. Distinguishing 
between feedback-regulated models, however, is difficult. In Figure~\ref{fig:beta.models} 
we compare several sub-classes: all follow a 
similar behavior and reflect the basic scalings in \citet{silkrees:msigma}. 
The models are derived in \citet{hopkins:seyferts} treating feedback-driven 
outflows as similarity solutions for expanding winds/blastwaves; the solutions 
adopt various equations of state and make different assumptions for the 
feedback coupling (whether e.g.\ feedback instantaneously couples to 
the ISM or continues to couple continuously throughout the event). Others \citep{menci:sam,
granato:sam} have obtained similar conclusions in different feedback models. 
We also compare the fits to the hydrodynamic simulations in \citet{hopkins:faint.slope} and 
\citet{younger:minor.mergers} -- the simulations give similar results for typical 
spheroids, but break the strict (analytically assumed) self-similarity: they predict a weak 
mass dependence (more massive galaxies being more gas poor and bulge-dominated, 
and having somewhat more violent resonant angular momentum transport and 
therefore more sharply peaked quasar/starburst activity as a 
consequence). The mass dependence amounts to a predicted 
$\beta^{\prime}\approx-0.25$ in Equation~\ref{eqn:beta.vs.mbh}. 
The observations favor the self-regulated class of models, and may weakly favor 
the mass dependence predicted in simulations. 

Some space of self-regulated models is ruled out. For example, 
\citet{hopkins:seyferts} argue that if accretion grows primarily out of cold instabilities 
``falling out of'' the expanding feedback-driven blastwave (in an adiabatic 
gas with Bondi-like spherical accretion), then very steep $\beta\sim31/19$ 
at early times is expected (only decaying to the $\beta\sim0.5-0.6$ 
seen in simulations in the late-time limit). Such a steep $\beta$ (shallow 
lightcurve decay) is not allowed.

\section{Constraints on AGN Lifetimes} 
\label{sec:compare.lifetimes}

These fits also allow us to quantify the quasar lifetime. 
There are, in fact, three different ``quasar lifetimes'' to which we could refer, 
which are constrained to varying degrees by the observations. 

{\bf (1) ``Effective'' Lifetimes}: This is the lifetime defined in terms of the duty 
cycle -- i.e.\ we define the effective quasar lifetime (at a given Eddington 
ratio $\mdot$ at a given BH mass $M_{\rm BH}$) by the duty cycle 
${\rm d}\delta(\mdot\,|\,M_{\rm BH})/{\rm d}\log{\lambda}$. 
Specifically, we invert Equation~\ref{eqn:duty.cycle.translation} to obtain
\begin{equation}
\frac{{\rm d}t_{\rm eff}(\lambda\,|\,M_{\rm BH})}{{\rm d}\log{\lambda}}\equiv
t_{H}(z)\,\frac{{\rm d}\delta}{{\rm d}\log{\lambda}}, 
\label{eqn:teff.defn}
\end{equation}
or in terms of the integrated duty cycle above $\lambda$ ($\delta[>\lambda]$), 
we have $t_{\rm eff}(>\lambda)\equiv t_{H}(z)\,\delta(>\lambda)$. 
This is directly determined by the observed Eddington ratio distributions 
at any redshift $z$, and is what is often referred to as the 
implied ``AGN lifetime'' from various observed statistics.\footnote{In
standard parlance, $t_{\rm eff}$ is {\em defined} in terms of the 
Hubble time $t_{H}$. However, strictly speaking $t_{\rm eff}$ has 
physical meaning (i.e.\ represents the time systems spend ``on'' at 
some $\lambda$ at a given redshift) only if two conditions are met: 
$t_{\rm eff}\ll t_{H}$ (the duty cycle $\delta\ll 1$) and 
host properties/triggering rates are varying on a timescale $\gg t_{\rm eff}$ 
(of order $t_{H}$). Both of these are satisfied at low redshift, for all 
but the lowest-$\lambda$ populations, but may not be true at high redshifts 
(imagine, for example, all hosts are ``just formed'' at some high redshift 
and are also all ``on'' -- then $t_{\rm eff}=t_{H}$, but the real lifetime 
can be significantly shorter).} 

{\bf (2) ``Integrated'' Lifetimes}: Recall, Equation~\ref{eqn:duty.cycle.translation} 
is only approximate. 
If one could view a BH or 
appropriate BH sub-population of final (remnant $z=0$) mass $M_{\rm BH}$ over 
its complete history, then the {\em true} ``integrated'' lifetime or duty cycle 
at a given Eddington ratio or luminosity would be given by the appropriate 
integral 
\begin{equation}
t_{\rm int}(>\lambda) = \int{\delta(>\lambda\,|\,z)}\,\frac{{\rm d}t}{{\rm d}z}\,{\rm d}z.
\label{eqn:tint.defn}
\end{equation}
It is immediately clear by comparison with Equation~\ref{eqn:teff.defn} that 
the integrated lifetime is equal to the effective lifetime if 
the duty cycle $\delta(>\lambda\,|\,z)$ is relatively constant over the 
redshift range of interest. In practice, so long as $\delta(>\lambda\,|\,z)$ has evolved 
relatively weakly between the time when most BHs of mass $M_{\rm BH}$ 
were formed and the observed redshift, one will obtain $t_{\rm eff}\approx t_{\rm int}$. 
This is seen to be the situation for relatively low-mass 
BHs corresponding to e.g.\ low-luminosity broad-line AGN \citep[which both 
are observed to have relatively constant number densities and duty cycles and 
are still growing, so their observed growth corresponds to that during their ``growth 
epoch''; see e.g.][and references therein]{ueda03:qlf,hasinger05:qlf,
hopkins:bol.qlf,shankar:bol.qlf}. For more massive BHs, however, where $\delta$ is 
significantly lower today than at high redshift, the two will not be the same. 

Nevertheless, the integrated lifetime is completely determined by the $z=0$ 
Eddington ratio distributions for most reasonable physical models. 
The reason is that this is an integral constraint, and the total accretion needs 
to sum appropriately to produce a $z=0$ BH of mass $M_{\rm BH}$. 
This is trivial to see if we temporarily consider the differential lifetime in terms of luminosity 
(rather than Eddington ratio) for a 
BH of final mass $M_{\rm BH}$: if we have some reasonable approximation
to the {\em shape} of the lifetime 
function (Equation~\ref{eqn:schechterfit.0}) -- then the total $M_{\rm BH}$ 
must be given by the integral: 
\begin{equation}
M_{\rm BH} = \int{L\,{\rm d}t} = \int{L\,\frac{{\rm d}t}{{\rm d}\log{L}}\,{\rm d}\log{L}} .
\label{eqn:continuity}
\end{equation}
In detail, we should truncate the lifetime function in Equation~\ref{eqn:schechterfit.0} 
at some minimum $L$ where the integrated lifetime $t\rightarrow t_{H}$ -- we do so 
but note 
in practice for the typical $\beta\approx0.6$ this lower 
$L$ is sufficiently small that the integral 
has already converged. 
If we assume $\beta(M_{\rm BH})$ and $\eta$ are relatively constant (and whatever 
physics sets them appears, at least in simulations, to be local to the AGN and 
redshift independent), then solving this equation yields a normalization $t_{0}$ 
appropriate for determining the {\em integrated} quasar lifetime above each $L$ 
\citep[for a more detailed derivation, see][]{hopkins:qso.all}. 

It is straightforward (although the continuity equations become somewhat more 
cumbersome and have to be solved numerically) 
to rewrite this derivation in terms of the Eddington ratio distribution 
(Equation~\ref{eqn:schechterfit}), and the solution is similar\footnote{In doing so, 
one is implicitly referring to the time spent at a given Eddington ratio 
{\em near the final BH mass} as observed at the appropriate redshift. 
There is no way to determine from the observations here, 
for example, if a $10^{9}\,\msun$ BH spent many Salpeter times 
growing in near-Eddington limited fashion from a small $\sim1\,\msun$ seed 
at very high redshifts, or formed directly as a $\sim10^{5}\,\msun$ seed BH. 
More rigorously we could write this as $t[> L/L_{\rm Edd}(M_{\rm BH,\,f})]$. 
However in terms of the Eddington ratio distribution that would be measured 
at any epoch in a relatively narrow range of BH mass, the two are identical.}.
Performing 
this exercise, we can approximate the integrated lifetimes 
at each logarithmic range in $\mdot$ by the same 
Equation~\ref{eqn:schechterfit} (with $\eta\approx0.4$), but with an ``integrated'' 
normalization $t_{0}$ 
whose constrained value can be roughly approximated 
as $1.26\,(1-1.8\,[\beta-0.6])\times10^{8}\,$yr (near a 
constant $\sim10^{8}\,$yr, i.e.\ 
$\approx2-3$ Salpeter times, appropriate 
for growing a BH by an order of magnitude in mass). Integrating 
Equation~\ref{eqn:schechterfit} from some minimum $\lambda$ to $\lambda=1$ 
with this normalization, we obtain the required integrated lifetime in each 
$\lambda$ range. 

{\bf (3) ``Episodic'' Lifetimes}: This is the lifetime of an individual AGN/quasar 
``event'' or some effective width in each peak in the quasar lightcurve. 
If AGN are excited to high Eddington ratio by some sort of trigger, then the 
time they spend in a given range of $L$ or $\mdot$ as a consequence of 
just that trigger is the appropriate episodic lifetime. The duty cycle at a given 
redshift is (to lowest order) the product of the triggering rate and 
episodic lifetime: for a fixed episodic lifetime, 
if the triggering rate is doubled, the effective lifetime doubles 
as well. Recall, it is the duty cycle (volume-averaged fraction) of BHs at a 
given Eddington ratio/luminosity that is observationally constrained by what 
we consider here. Therefore, based on the observations discussed thus far, we 
can only place an upper limit on the episodic lifetime (it cannot, of course, be 
longer than the integrated lifetime). We discuss the degeneracies this implies 
and possibilities for direct constraints and complimentary lower limits to 
the episodic lifetime in \S~\ref{sec:lightcurves}. 

\begin{figure}
    \centering
    \scaleup
    \plotter{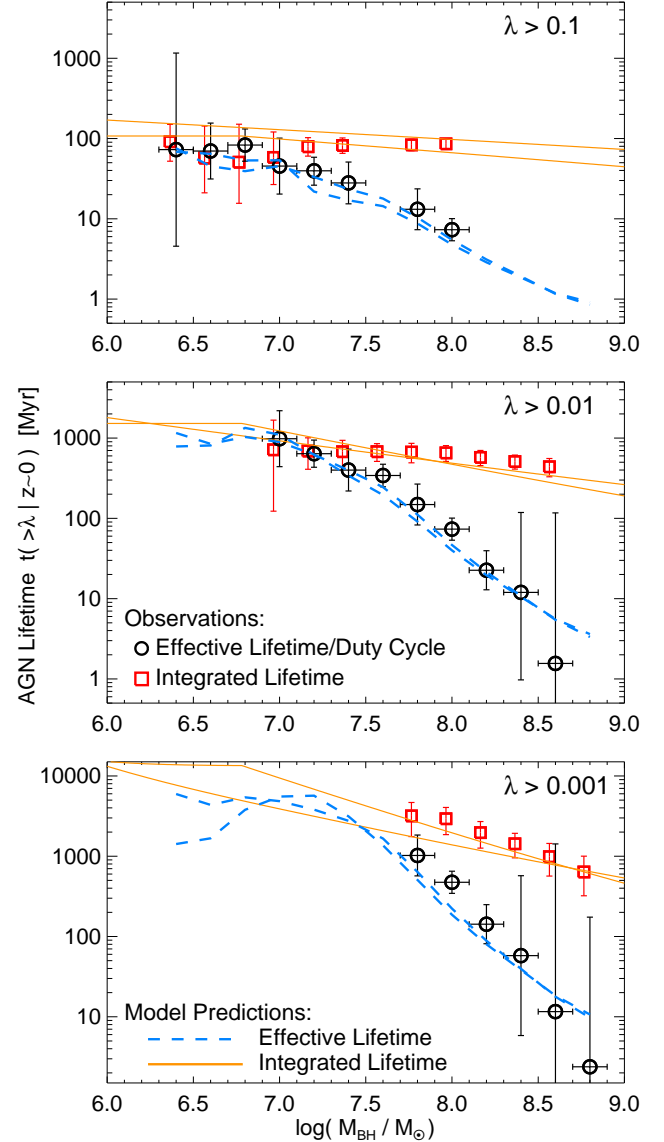}
    \caption{Effective quasar lifetime (black circles) above a given 
    Eddington ratio $\lambda$, i.e.\ the duty cycle $\delta(>\lambda)$ times 
    the Hubble time, fitted as a function of BH mass to the data in 
    Figures~\ref{fig:yulu.1}-\ref{fig:yulu.2}. The data are at low redshift, so this 
    should be thought of as the average lifetime for objects whose luminosity 
    function is similar to that observed at low-$z$ (see text). 
    Dashed blue lines compare the 
    model predictions, integrating over the cosmological 
    history of triggering. 
    The integrated quasar lifetime (red squares) determined by integrating over the 
    best-fit lifetime distribution to the observations in each bin of $M_{\rm BH}$ 
    (applying the appropriate mass conservation equation) is shown for 
    comparison. 
    Dotted orange lines are the predicted integrated quasar lifetimes -- reflecting the total time 
    at each $\lambda$ (integrated over any arbitrary cosmological history) 
    required by continuity to produce a BH at $z=0$ with the observed mass. 
    The two lifetimes are similar 
    for BHs at lower masses $M_{\rm BH}\lesssim10^{7.5}\,\msun$ (at which masses 
    observations suggest the duty cycle/volume density of quasars has been relatively 
    constant since $z\sim1-2$), but different at the highest masses 
    (where there has been a steep drop-off in quasar activity since $z\sim2$; 
    i.e.\ the systems did most of their growth preferentially at high redshifts and 
    have lower average integrated lifetimes at higher $\lambda$ today). 
    \label{fig:t0}}
\end{figure}

Figure~\ref{fig:t0} compares the effective and integrated lifetimes determined 
from our fitting as a function of BH mass and minimum $\lambda$. 
The effective lifetime, being essentially a duty cycle, can be determined 
directly from the Eddington ratio distributions where they cover the necessary 
dynamic range. The results are as expected -- roughly constant at 
low BH masses, with a steep fall in duty cycles/effective lifetimes 
at high-$M_{\rm BH}$. The integrated lifetimes are determined 
from the fitted lifetime distributions (Equation~\ref{eqn:schechterfit}), 
given the continuity requirement. They are also as expected -- 
for $\lambda\gtrsim0.1$, for example, this gives an expectation that 
most BHs spent $\sim$ a few Salpeter times at high Eddington ratio, 
a requirement in any self-consistent Eddington-limited model. They are 
much more weakly mass-dependent -- since 
$\mdot$ is dimensionless and BH growth at fixed $\mdot$ exponential, 
we would expect that (to lowest order) systems of different masses should 
have spent similar time at $\mdot$ over the course of their evolution. 
At high $\mdot$, this is true by definition; as noted above the time at a given 
$\mdot$ refers to time near that $\mdot$ near the final BH mass -- 
if the Eddington limit is applicable, all systems must spend a similar time in 
this regime.

Comparing the integrated and effective lifetimes yields an 
effective ratio of duty cycles: 
\begin{equation}
\frac{t_{\rm eff}}{t_{\rm int}} = \frac{\delta(>\lambda\,|\,z_{\rm obs})}
{\langle \delta(>\lambda\,|\,z>z_{\rm obs}) \rangle}, 
\label{eqn:duty.cycle.ratio}
\end{equation}
where $\langle \delta(>\lambda\,|\,z>z_{\rm obs}) \rangle$ is an 
appropriate weighted average $\delta$ (weighting over the 
time integral in Equation~\ref{eqn:tint.defn} and by the relative fraction of 
systems with mass $M_{\rm BH}$ at $z_{\rm obs}$ formed around 
each $z$). Roughly speaking, this yields the ratio of the duty cycle 
at the observed redshift $z_{\rm obs}$ to that during the epoch in which 
most BHs of the given mass were actively growing. For local low-mass 
BHs, this ratio is near unity; for high-mass BHs, is falls steeply to somewhere 
between $\sim1-10\%$ of its high-redshift value, consistent with 
the now-standard picture of ``downsizing'' in BH growth.

Note that the lifetimes in Figure~\ref{fig:t0} are not directly analogous to those 
in previous works \citep[e.g.][]{yutremaine:bhmf,salucci:bhmf,shankar:bhmf,marconi:bhmf}, 
because those assume a simplified light-bulb model, rather than a continuous distribution 
of accretion rates. 

Also, these models usually quote the total time 
at high Eddington ratio at all BH masses since some adopted initial condition 
(not the time at some Eddington ratio in a narrow range of BH mass, constrained 
directly by observations here); as such, the comparison requires matching the 
boundary conditions for seed BHs in the light-bulb models.\footnote{Specifically, 
we show in Figure~\ref{fig:t0} the average time spend in a given $\mdot$ 
range for individual BHs with a given $z=0$ mass. Some other 
definitions adopted in the literature reflect the {\em total} time for 
all BHs that are in or passed through a given BH mass interval 
(e.g.\ massive $z=0$ BHs when they were, earlier, at this mass). 
The latter lifetime definition will be higher at lower masses, 
since it includes the time local, more massive systems spent ``on'' 
at earlier times getting to their present-day masses.}
Also, $\delta$ here is the duty cycle above some $\lambda$, and is not 
directly comparable to the ``AGN fraction'' often defined observationally. 

For example, \citet{heckman:local.mbh} point out (as seen here in the same data) 
that $\delta(>\lambda)$ decreases strongly with $M_{\rm BH}$; however, 
\citet{kauffmann:qso.hosts} and \citet{kewley:agn.host.sf} argue from the 
same data that the ``AGN fraction'' increases with galaxy mass. 
This apparent contradiction owes to a number of well-known trends: bulge-to-disk ratios 
and velocity dispersions drop rapidly in low-mass galaxies, so the span of 
BH mass and galaxy mass are not the same at low masses; specific star formation 
rates also increase rapidly in low-mass galaxies, so for the same fractional Eddington 
luminosity low-mass systems will be less AGN-dominated; and most such samples 
are AGN luminosity-limited, and so probe more massive hosts to lower 
Eddington ratios. Modeling these selection effects is non-trivial and quite sensitive 
to the specific selection effects and wavelengths of each observed sample -- we 
therefore do not attempt a direct comparison here. We do note, however, that 
the comparison of the \citet{heckman:local.mbh,kauffmann:new.mdot.dist} and 
\citet{kauffmann:qso.hosts,kewley:agn.host.sf} samples indicates the consistency 
of these different indicators; in \citet{hopkins:groups.qso} the authors attempt to 
model similar selection effects to compare the self-regulated 
lightcurves considered here with AGN fractions in the samples observed 
by \citet{kauffmann:qso.hosts} at low redshift and \citet{erb:lbg.gasmasses,
kriek:qso.frac} at high redshift 
\citep[see also][and references therein]{silverman:qso.hosts}, and find consistent results.

\section{Translation to AGN Lightcurves: Episodic versus Integrated Lifetimes}
\label{sec:lightcurves}

Given the quasar lifetime distribution, we can ask how this relates to 
some ``average'' quasar lightcurve $L(t\, | \, M_{\rm BH})$. 
{\em If the average lightcurve were monotonic}, this would be trivial: 
inverting $\dtdlogL$ and integrating defines $L(t)$. 
For the Schechter function in Equation~\ref{eqn:schechterfit} this is numerically 
straightforward but tedious; we find a statistically identical 
answer with the convenient analytic representation of the lightcurve 
\begin{equation}
\mdot = {\Bigl[}1 + (t/t_{Q})^{1/2} {\Bigr]}^{-2/\beta}
\label{eqn:lightcurve}
\end{equation}
where $\beta$ is the same fitted to the lifetime distribution in Equation~\ref{eqn:schechterfit} 
and 
\begin{equation}
t_{Q} = t_{Q}^{\rm max} \equiv t_{0}\,\frac{\eta^{\beta}}{\beta\,\ln{10}}
\label{eqn:lightcurve.tQ}
\end{equation}
relative to the $t_{0}$ fitted in Equation~\ref{eqn:schechterfit} (the zero-point in $t$ 
in Equation~\ref{eqn:lightcurve} is obviously arbitrary and here is fixed to 
the time when the lightcurve is at maximum). For the 
median $\beta\approx0.6$, this gives $t_{Q}\approx5.3\times10^{7}\,$yr, 
similar to the Salpeter time and characteristic dynamical times in the 
central regions of galaxies. 
It is straightforward to see that the lifetime distribution yielded by this lightcurve is 
statistically a good match to the Schechter function fits and observations in 
\S~\ref{sec:compare.dist}, and it is convenient for analytic models of quasar evolution. 
If $t_{Q}\rightarrow t_{Q}/2$ and $t\rightarrow | t |$, it is trivial to treat this as 
time-symmetric about some peak, and a nearly identical 
lifetime distribution is obtained if one assumes exponential growth up 
to a peak luminosity followed by a ``decay phase'' given by 
Equation~\ref{eqn:lightcurve}. 

Equation~\ref{eqn:lightcurve} is a good match to the decline or 
blowout phase of quasar evolution in feedback-regulated 
simulations \citep{hopkins:faint.slope}. The assumption of monotonicity 
need not be exact, 
provided that the (average) fractional amplitude of variation in the quasar 
luminosity is small (factor $\lesssim 2-3$) on timescales shorter than the 
integrated quasar lifetime at the luminosity of interest. At some point 
as the quasar lifetime approaches the Hubble time this is almost certainly not 
true; however, for the short, high-$L$ periods 
this is reasonable.

\begin{figure}
    \centering
    \scaleup
    \plotone{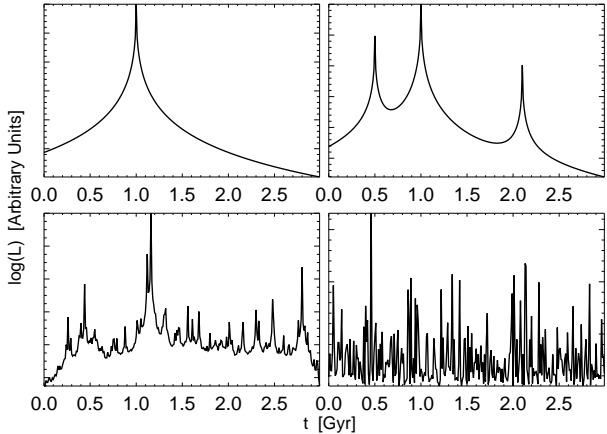}
    \caption{Examples of different quasar lightcurves that match the observed 
    quasar lifetime distribution (Equation~\ref{eqn:schechterfit}). The models are distinguished
    by different {\em episodic} quasar lifetimes and frequency of triggering. 
    The integrated quasar lifetime/duty cycle and average post-peak decay 
    are the same (required to match the observed $\mdot$ distributions) -- the events decay 
    after each peak following Equation~\ref{eqn:lightcurve}. 
    Different observations, for example probing the transverse proximity effect, are required 
    to break the degeneracies between these models. Constraints from these observations 
    at present appear to rule out the model in the lower right panel, but the others, with an 
    episodic lifetime $\sim0.3-1$ times the effective/integrated lifetime, are allowed. 
    \label{fig:lightcurves}}
\end{figure}

What if, however, there were two such episodes (each compressed in 
duration and separated by a cosmological interval $\gtrsim$Gyr)? 
Figure~\ref{fig:lightcurves} shows a few such 
models (arbitrarily re-normalized to produce the same 
final BH mass).  First, a monotonic or single event case (for convenience, we assume the 
rise and fall about each ``event'' are symmetric). 
Second, a superposition of three such events, each with shorter duration 
so that the sum $\dtdlogl$ is conserved.  
Third, a superposition of a couple of bright events and a larger number of lower-peak luminosity 
events, with a power spectrum 
following $\dtdlogl$.  And fourth, a superposition of such events with effectively short episodic 
lifetimes ($t_{Q}\ll t_{\rm eff}$), or equivalently large-amplitude variability 
on short timescales added according to a power-law spectrum with slope $\beta$. 

All of these model lightcurves reproduce the same lifetime distribution $\dtdlogl$, and 
given $\dtdlogl$ and the observed Eddington ratio distribution, the observed 
QLF will necessarily be the same. Moreover, they can all be normalized to 
give the same final BH mass (although for this final BH mass, they may correspond 
to different initial BH masses; but this is much smaller than the final mass 
and so not constrained by the observations considered). 
Measurements of the Eddington ratio distribution cannot 
uniquely determine the average lightcurve; only $\dtdlogl$. 

In physical and observable terms, this is because the Eddington ratio distribution and 
corresponding $\dtdlogl$ yield only the {\em effective} and {\em integrated} quasar lifetimes 
-- i.e.\ the duty cycle and total lifetime -- 
at some $L$, not the {\em episodic} lifetime. Averaged over some redshift 
interval, the time $t(L)$ determined from 
$\dtdlogl(L)$ is the {\em total} time that an object will be active at that $L$ in the 
interval. Whether that time came continuously -- in a single episode -- or in some 
large number of shorter episodes, cannot be inferred from the observations 
considered thus far. 

The episodic lifetime is of particular 
interest because it relates to the triggering 
rate of quasars. If quasars are triggered at a rate $\dot{n}$ 
and have an episodic lifetime (i.e.\ a duration around luminosity $L$ per peak or 
triggering episode) of $t_{i}$ ($\sim t_{Q}$ if the event is similar to the description of 
Equation~\ref{eqn:lightcurve}), then the number density observed is 
$n=\dot{n}\,t_{i}$. If we know the observed number density and the effective lifetime $t_{\rm eff}$, 
then attempting to infer the triggering rate we can only obtain
\begin{equation}
\dot{n} \approx \frac{n}{t_{\rm eff}}\,{\Bigl(}\frac{t_{\rm eff}}{t_{i}}{\Bigr)}, 
\end{equation}
i.e.\ we are limited by some guess for the ratio $t_{i}/t_{\rm eff}$. 
The statement that $t_{i}=t_{\rm eff}$ is equivalent to the assumption that every 
object at a given $L$ has had only one triggering episode in recent time. 
But if $t_{i}$ were smaller than $t_{\rm eff}$ by some ratio $N$ ($t_{i}=t_{\rm eff}/N$), then 
the implied triggering rate for the same number of observed systems 
is higher by a factor $N$. 

Assuming a physical model for what triggers quasars, we can use observations of 
quantities such as e.g.\ the galaxy merger rate to determine $\dot{n}$, 
and doing so for the case of galaxy mergers gives the result that 
$t_{i}\sim t_{\rm eff}$ (i.e.\ there are typically $\sim1$ triggers per unit Hubble time). If 
we wish to determine $\dot{n}$ 
without reference to a specific quasar fueling model, then we need 
the episodic lifetime $t_{i}$. 

The observed rate of 
evolution of the quasar luminosity function places an upper limit $t_{i}\lesssim 10^{9}\,$yr 
at high luminosities \citep{martini04}, but this is already larger than the integrated 
lifetime at these luminosities. Lower limits can be derived from the sizes of narrow 
line regions \citep{bennert:nlr.structure}, but these are short 
($t_{i}>3\times 10^{4}\,$yr), so are not especially constraining 
($t_{i}\ll t_{\rm eff}$ is unlikely, as it requires extremely high rates of triggering, on 
timescales faster than the relevant dynamical times). Indirect 
measures relying on matching 
the observed BH mass function and QLF 
\citep{yutremaine:bhmf,salucci:bhmf,shankar:bhmf,marconi:bhmf} 
and halo occupation models which compare 
clustering data \citep{porciani2004,croom:clustering,
fine:mbh-mhalo.clustering,porciani:clustering,daangela:clustering} 
constrain only the effective lifetimes. 

One probe of episodic lifetimes is the 
transverse proximity effect -- 
if individual episodes are short, the sizes of 
ionized bubbles around quasars should be smaller
than if individual episodes are long.
Preliminary constraints from non-detections  
\citep[see e.g.][for a review]{schirber:transverse.proximity,martini04} suggest 
lifetimes $t_{i} > 10^{7}\,$yr (with a strong limit $t_{i}\gg 10^{6}\,$yr)
at high luminosities ($\lambda\gtrsim0.1$) where the observations are possible. More recently, 
potential detections \citep{jakobsen:heII.ion.transverse.proximity,
goncalves:transverse.proximity} and indirect proximity effects 
\citep{worseck06:indirect.transverse.proximity,worseck07:indirect.transverse.proximity} 
in a few objects suggest 
$t_{i}\approx 2.5-5\times10^{7}\,$yr
(almost exactly $t_{i}=t_{\rm eff}\sim t_{Q}^{\rm max}$ for $\mdot \gtrsim 0.1-0.2$). 
Similar analysis of the Gunn-Peterson effect around high-redshift ($z>6$) 
quasars suggests comparable episodic 
lifetimes \citep{bajtlik:gunnpetersen.qso.age.est,haiman:gunnpetersen.qso.age.est,
yulu:stromgren}, but is more subject to uncertainties in the structure of the 
surrounding gas \citep{lidz:proximity}. 

The lengths of relativistic jets and radio lobes also imply episodic lifetimes, but it is 
not clear that the lifetime of radio loud activity or large-scale jet formation is 
the same as (or even correlated with) the lifetime for specific bolometrically 
luminous activity. These observations 
\citep[see e.g.][]{scheuer:radio.jet.size.lifetimes,
blundell:radio.jet.size.lifetimes} suggest lower-limits for 
$t_{i}\gtrsim$ a few $10^{8}\,$yr, but it is important to note that the observed 
systems with large jets in these samples are primarily at lower Eddington ratio 
$\sim0.01$, so this is comparable to or a factor of a few smaller than the total lifetime 
at these luminosities. For more luminous FR II sources, a lifetime similar 
to that from the transverse proximity effect, $\approx 2\times10^{7}\,$yr, 
has been estimated \citep{bird:bright.jet.lifetimes} \citep[although see 
also][]{reynolds.begelman:radio.source.sizes,merloniheinz.bhfp.radio}.

These observations, although tentative, seem to suggest an episodic lifetime 
similar to the total lifetime, in the range 
$t_{i}/t_{\rm eff}\approx 0.3 - 1$ over reasonably high luminosity ranges $\mdot\sim0.01-1$. 
This is consistent with simulations and cosmological models, and similar to the 
characteristic timescales of the problem 
(the Salpeter time for the growth of the BH, $4.2\times10^{7}\,$yr, 
and characteristic dynamical/free-fall times in galaxy centers, 
$\sim10^{7}-10^{8}\,$\,yr). Improved constraints would permit the 
extension of this comparison to lower Eddington ratios and enable observations to 
uniquely determine the {\em triggering rates} 
of AGN as a function of BH mass, luminosity, and redshift.

\section{Redshift Evolution: Quenching and Downsizing}
\label{sec:z.evol}

Most of the observations considered here are at low redshift. 
Some redshift evolution 
in Eddington ratio distributions {\em must} occur, as indicated in Figure~\ref{fig:t0}: 
$t_{\rm eff}\ll t_{\rm int}$ for massive BHs at low redshift, so there must have been 
some point at higher redshift where the duty-cycle of high-mass BHs was greater 
(when they accreted most of their mass, around $z\sim2$). 
Constraints on this evolution can be obtained from the evolution of 
the quasar luminosity function -- 
however, as discussed above, the QLF is primarily sensitive (especially 
at high redshifts, where the faint end is less well-constrained) to high 
luminosities $\lambda\gtrsim0.1$; likewise integral constraints from the 
shape of the BH mass function (and the requirement that the integrated 
luminosity yield the appropriate final BH mass) primarily relate to 
these Eddington ratios, where most mass is gained. 

As a consequence, the QLF constrains only the appropriate combination 
of $\beta$, $t_{0}$, and $\eta$ in Equation~\ref{eqn:schechterfit} such that 
the duty cycle at high Eddington ratio is reproduced. 
As noted earlier, the duty cycle is equivalent to the effective lifetime 
$t_{\rm eff}$ ($t_{\rm eff}(\lambda>0.1)/t_{H}(z)$ being the 
high-$\mdot$ duty cycle). If the lifetime/duty cycle is parameterized 
in the form of Equation~\ref{eqn:schechterfit}, then it is straightforward to 
show that this duty cycle of interest is 
\begin{equation}
t_{\rm eff}(\lambda>0.1) \approx t_{0}\,\eta\,\exp{(-0.1/\eta)}, 
\end{equation}
essentially independent of the slope parameter $\beta$ (that 
affecting the lifetime only at lower $\mdot$) and only weakly 
dependent on $\eta$; for the reasonable physical range 
$\eta\sim0.4-1$, $t_{\rm eff}$ changes by a factor $\sim2-3$ 
($\eta$ cannot decrease much below this with 
redshift, or the number density of bright objects, observed to rise, 
would be exponentially suppressed). 
Various observations and synthesis models have been 
used to constrain this duty cycle -- since this is just 
the high-$\mdot$ duty cycle, ``light bulb'' and more sophisticated models 
yield nearly identical results here \citep[see e.g.][]{haehnelt:bh.synthesis.model,yutremaine:bhmf,yulu:bhmf,
haiman:bhmf,marconi:bhmf,shankar:bhmf}. 
From the fits of these synthesis models to the QLF, or from 
constructing an analogous simple model (assuming the lightcurves here, 
but fitting to the observed QLF from \citet{hopkins:bol.qlf} and 
integrating the continuity equations as in the synthesis models above), 
it can be seen that the duty cycle at fixed BH mass increases with 
redshift (at least from $z\sim0-2$) in approximate power-law fashion, 
i.e.\ as 
\begin{equation}
t_{\rm eff}(\lambda>0.1)  \propto (1+z)^{\alpha}
\label{eqn:z.evol}
\end{equation}
with a maximum $t_{\rm eff}=t_{\rm int}\approx 10^{8}\,$yr for this $\lambda$ (obviously 
$t_{\rm eff}$ cannot increase beyond this point -- and indeed 
this reproduces the ``flattening'' in duty cycles at $z\gtrsim2$).

Clearly, the duty cycle of high-mass systems must increase 
more quickly with redshift than that of low-mass systems. 
From the comparison with observations in the various synthesis models above, 
we note that the needed evolution can be crudely approximated as 
\begin{equation}
\alpha = \ln{{\Bigl[}1 + \frac{M_{\rm BH}}{10^{7}\,\msun} {\Bigr]}}\ .
\label{eqn:z.evol.alpha}
\end{equation}
This is not a rigorous derivation; it simply provides a useful interpolation 
formula that approximately reproduces the QLF evolution. 
At lowest order, this power-law increase in the duty cycle of 
high-$\mdot$ activity simply reflects the evolution in the observed number 
density of quasars at a given $L=L_{\rm Edd}(M_{\rm BH})$ 
\citep[see e.g.][]{hasinger05:qlf}. Second order corrections come from e.g.\ the 
evolving number density of BHs, but these depend only weakly on the 
lightcurve model in this $\mdot$ range. 

Given this, the simplest possible model is indeed consistent 
with the observations: a model in which quasar lightcurves 
($\beta$ and $t_{Q}$, in Equation~\ref{eqn:lightcurve}) are 
redshift-independent. Effectively this means $\beta$ and $\eta$ in Equation~\ref{eqn:schechterfit} 
are fixed, as is the episodic lifetime of quasars -- all physics local to AGN 
evolution are redshift-independent, only the triggering rate (hence the 
duty cycle and corresponding $t_{\rm eff}$ and $t_{0}$ in Equation~\ref{eqn:schechterfit}) 
evolve. This is demonstrated in models that adopt this assumption 
\citep{hopkins:qso.all,hopkins:bol.qlf,hopkins:groups.qso,
hopkins:groups.ell}, but the form of the constraints in Equation~\ref{eqn:z.evol} 
makes it implicit. 

It is clearly important to obtain 
direct constraints on the Eddington ratio distributions 
at high redshift, particularly constraints on the low-$\lambda$ population, 
in order to break the degeneracies between this redshift-independent model 
and one in which quasar lightcurves evolve. If there is some evolution, 
it may indicate a difference in fueling or feedback modes: one could imagine 
different lightcurves resulting if the bright, high-redshift 
population is fueled by violent mergers where the low-redshift population 
is fueled by stochastic mechanisms, bar-induced inflows, or minor 
mergers; or if feedback is not important in some populations -- at high redshift, 
feedback may act efficiently as the systems of interest are more bulge-dominated 
(leading to more sharply peaked lightcurves), whereas at low redshift, with 
activity in primarily low-$M_{\rm BH}$ systems with large, gas-rich disks, feedback 
may be ineffective at expelling gas content and suppressing inflows on 
large scales.

\begin{figure}
    \centering
    \scaleup
    \plotter{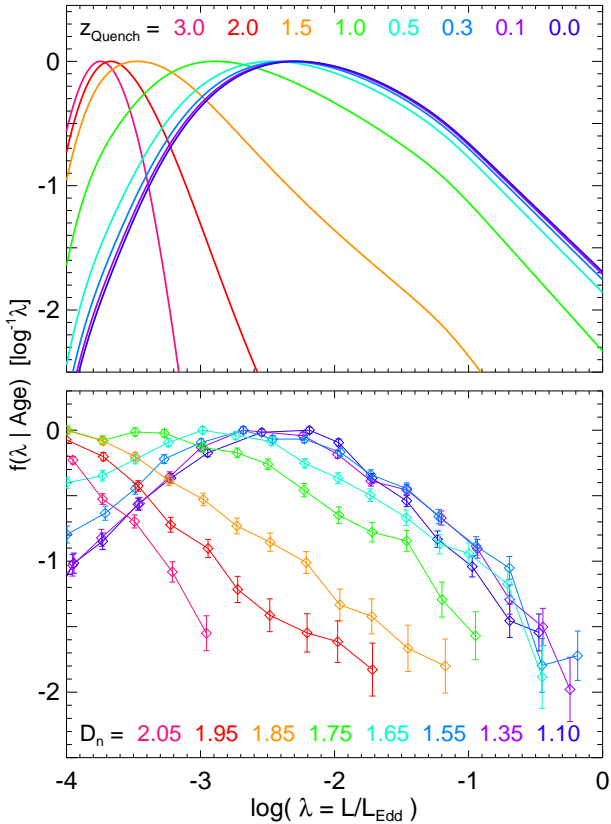}
    \caption{{\em Top:} Distribution of Eddington ratios (arbitrary units) predicted 
    for populations with the same lightcurve (Equation~\ref{eqn:lightcurve}, 
    with $\beta=0.6$) but different toy model triggering histories. Triggering 
    rises with time in the same manner until a redshift $z_{\rm Quench}$, when 
    the systems are shut down (a roughly Gaussian rise/fall motivated by observations). 
    For all still-active/young systems (low $z_{\rm Quench}$), the distributions asymptote to a 
    single distribution given by the ratio of $t(>L)/t_{H}$ for a single lightcurve/event 
    (Equations~\ref{eqn:dtdlogl.defn}-\ref{eqn:duty.cycle.translation}; 
    with a turnover once $t(>L)\sim t_{H}$). 
    For quenched/inactive systems (high $z_{\rm Quench}$), the distributions ``pile up'' 
    in the long power-law tail of decayed luminosities since the epoch of triggering. 
    {\em Bottom:} Observed distributions from \citet{kauffmann:new.mdot.dist} at 
    fixed BH mass ($\sim10^{7}-10^{8}\,\msun$) as a function of stellar population 
    age (the parameter $D_{n}$; $D_{n}\lesssim 1.5-1.6$ corresponding to star-forming/still 
    active systems, $D_{n}\sim2$ corresponding to ``red and dead'' systems with 
    stellar population ages $\sim t_{H}$). The observed distribution reflects the predicted 
    trend: a single (redshift-independent) lightcurve is consistent with the observed 
    dependence of $\mdot$ on age/star-forming classification. The ``log-normal'' behavior 
    in young populations simply reflects the universal nature of the lightcurve and 
    inevitable turnover when $t_{\rm eff}(>\lambda)\sim t_{H}$; allowed to 
    decay after quenching, this becomes the power-law-like tail in old populations.  
    \label{fig:mdot.vs.age}}
\end{figure}

There are some constraints on $\beta$ and $t_{Q}$ that can be obtained 
from low-redshift observations. As discussed above, if a population is 
still growing (i.e.\ there is no sharp feature in the redshift history of 
triggering/growth) then the $\mdot$ distribution today simply reflects 
typical low-redshift lightcurves. However, if there are distinct differences in the 
redshift history of triggering -- if one population (at the same BH mass) 
is ``quenched'' (ceases growth/new triggering) at a different time than another -- 
then the resulting Eddington ratio distribution (for the same quasar lightcurves) 
will not be the same at $z=0$. 

Figure~\ref{fig:mdot.vs.age} demonstrates this 
with a very simple toy model. Assume that AGN lightcurves are universal and 
redshift-independent, given 
by Equation~\ref{eqn:lightcurve} with $t_{Q}=5\times10^{7}\,$yr and $\beta=0.6$. 
Objects have luminosity $L=0$ 
until ``triggered'' at some time, then follow this simple lightcurve decay 
($t$ in Equation~\ref{eqn:lightcurve} is the time since the trigger). 
The probability of a ``trigger'' as a function of time we arbitrarily 
parameterize as a Gaussian in cosmic time, rising from high redshift 
($z\sim6$) to some peak 
at the ``quenching redshift'' $z_{\rm Quench}$ and then declining to $z=0$ 
Specifically, 
\begin{equation}
P({\rm trigger}\,|\,z) \propto \exp{{\Bigl\{} 
\frac{t_{H}(z)}{t_{H}(z=6)} - {\Bigl [}\frac{t_{H}(z)}{t_{H}(z_{\rm Quench})} {\Bigr ]}^{2} {\Bigr\}}}\ .
\end{equation}
This is roughly chosen to correspond to the shape of the evolution in the 
observed AGN luminosity density, but we emphasize that it 
is just for illustrative purposes. 
Figure~\ref{fig:mdot.vs.age} shows the resulting $z=0$ distribution in 
$\mdot$, for a set of Monte Carlo populations evolved in this toy model, but 
with a different quenching redshift $z_{\rm Quench}$ for each. 

Populations that quench early (high $z_{\rm Quench}$) have 
all decayed for a long time, to pile up at low $\mdot$ and form a 
power-law distribution in $\mdot$ above some very low minimum $\mdot$. 
(The power-law {\em must} turn over at some sufficiently low $\mdot$, 
such that the duty cycle integrated over all Eddington ratios is unity; 
the specific turnover around $\mdot\sim10^{-4}$ in Figure~\ref{fig:mdot.vs.age} 
reflects the specific lightcurve and triggering functional forms assumed 
here, together with the age of the Universe. Adjusting the late-time lightcurve behavior 
or early-time triggering rates can shift this to lower $\mdot$.)

Populations that 
quench late (low $z_{\rm Quench}$) -- i.e.\ approach the limit of continuous growth 
still occurring today -- show a power-law behavior at high-$\lambda$ 
that directly traces the quasar lightcurve (as discussed above, with duty cycles 
simply given by $t_{\rm eff}\sim t_{\rm int}$), with a turnover where 
$t_{\rm eff}\rightarrow t_{H}$. The resulting distribution looks roughly log-normal, 
and asymptotes to the same distribution for all low-$z_{\rm Quench}$ (all still-active 
populations). 

Note that the details in Figure~\ref{fig:mdot.vs.age} are somewhat arbitrary -- 
we have just chosen a toy model triggering history to highlight the 
dependence on models with strong features in that history; it is easy to construct others. 
It is possible to tune these histories such that, for example, the high-$z_{\rm Quench}$ 
predictions continue their power-law like behavior to lower $\mdot$ or 
shift the ``turnover'' in the lognormal (low-$z_{\rm Quench}$) 
regime. But the {\em qualitative} results are insensitive to the precise 
parameterization of the triggering history, provided there is a significant feature/shutdown 
after a given time. 

Recently, \citet{kauffmann:new.mdot.dist} expanded upon the Eddington ratio distributions 
measured by \citet{heckman:local.mbh} and \citet{yu:mdot.dist}, and quantified 
the $\lambda$ distribution as a function of stellar population age (specifically the 
observable parameter $D_{n}$, which is a tracer of activity: $D_{n}\lesssim1.5$ 
systems being still active or $\lesssim 3\,$Gyr old, in mean stellar population age, 
whereas $D_{n}\rightarrow2$ systems are quenched with ages $\sim t_{H}$). 
Figure~\ref{fig:mdot.vs.age} compares the $\lambda$ distribution measured 
as a function of stellar population properties (for BHs with fixed mass $10^{7}-10^{8}\,\msun$) 
at $z=0$. The trends are very similar to those predicted (and it is not hard to 
imagine more detailed star formation history models, 
more precisely tuned to yield quantitative agreement). 

This is already interesting: \citet{kauffmann:new.mdot.dist} interpret the observed 
trend as an indicator of two independent accretion modes (a ``power-law'' mode, 
in old systems, and a ``log-normal'' mode, in young ones). This could still be the case, 
but the comparison here demonstrates that the observed trend is also the natural 
expectation of a model in which all AGN lightcurves are identical, but there 
are simply differences in the triggering rate corresponding to when different galaxies/BH 
populations were ``quenched'' or slowed their growth. The uniformity of the Eddington 
ratio distribution in the ``log-normal'' regime is not surprising in this case -- it simply 
reflects the fact that quasar lightcurves are similar and that all populations in this 
regime of stellar population age are ``still active'' (i.e.\ are continuously growing, in a 
population-averaged sense; they do not have a sharp feature of higher-redshift activity 
that dominated their growth); the ``power-law'' regime is simply the ``log-normal'' population 
allowed to decay to lower luminosities as triggering rates decline with redshift.

\begin{figure}
    \centering
    \scaleup
    \plotone{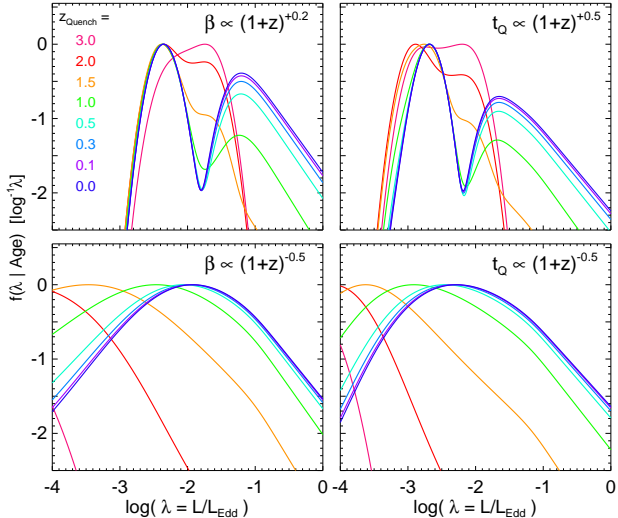}
    \caption{As Figure~\ref{fig:mdot.vs.age}, but in models where the median lightcurve 
    (parameterized by lifetime slope $\beta$ and episodic lifetime $t_{Q}$; 
    Equation~\ref{eqn:lightcurve}) evolves with redshift. 
    {\em Top:} Lightcurves become more shallow/extended with increasing 
    redshift ($\beta\propto(1+z)^{0.2}$, $t_{Q}\propto(1+z)^{0.5}$). 
    It is not possible to match the observed trends from 
    Figure~\ref{fig:mdot.vs.age} with such evolution. 
    {\em Bottom:} Lightcurves become more sharply peaked with increasing 
    redshift ($\beta\propto(1+z)^{-0.5}$, $t_{Q}\propto(1+z)^{-0.5}$). 
    Such evolution is consistent with the observed trends (and $z\sim1$ observations 
    in Figure~\ref{fig:merloni}). The trend of $\lambda$ distribution with 
    age/triggering history puts significant constraints on lightcurve evolution: 
    evolution towards more ``quiescent'' models (including stellar wind or isolated accretion 
    disk modes) is disallowed; no evolution or weak evolution towards more 
    violent/efficient feedback modes are consistent with observations. 
    \label{fig:mdot.vs.age.evol}}
\end{figure}

Moreover, these observed trends place significant constraints on lightcurve 
evolution. Figure~\ref{fig:mdot.vs.age.evol} considers the same model predictions, 
but in models where the lightcurve 
parameters in Equation~\ref{eqn:lightcurve} evolve with the 
``triggering redshift.'' For 
convenience we parameterize the evolution simply as 
$\beta \propto (1+z)^{\beta^{\prime}_{z}}$ and $t_{Q} \propto (1 + z)^{t^{\prime}_{z}}$, 
where $z$ here refers to the redshift where a given trigger occurs. 
For $\beta^{\prime}_{z}=-0.5$ or $t^{\prime}_{z}=-0.5$, the predictions are similar 
to the no-evolution case, consistent with the observations. 
Again, we stress that the exact mapping to observations depends on the precise 
redshift evolution of the triggering distribution, and more importantly 
on how star formation histories in detail evolve relative to AGN triggering 
histories, needed to predict observed quantities such as $D_{n}$ -- however, 
as long as the observed qualitative features are intact, it is always possible to 
find models where those assumptions are adjusted to give a more 
precise quantitative match to the observations. 
However, for $\beta^{\prime}_{z}=0.2$ or $t^{\prime}_{z}=0.5$, 
the results are qualitatively much different -- they do not resemble the observations. 
The sense of the implied evolution in this case would be that lightcurves 
become more shallow/extended with 
redshift: as a consequence, quenched systems would not decay sufficiently 
relative to observations. In these cases, where the features of the predicted distributions 
are qualitatively different from the observations, we find that 
no tuning of the triggering rate distributions is able to ``fix'' the disagreement 
with observations. 

This comparison constrains the allowed range: 
\begin{eqnarray}
\nonumber - 0.5\, &\le \beta^{\prime}_{z} \le& 0.05 \\ 
- 0.7\, &\le t^{\prime}_{z} \le& 0.25\ .
\label{eqn:z.evol.constraints}
\end{eqnarray}
These are not error bars; rather, they represent the allowed 
range in which solutions exist that can reproduce 
the observed trends from \citet[][]{kauffmann:new.mdot.dist}. 
Within this range, the results are also consistent with the 
observationally inferred Eddington ratio distributions from 
\citet{merloni:mdot.dist.fits.prep} at $z=1$. 
The constraints are non-trivial: redshift evolution must be 
relatively mild, and the form of evolution allowed is such that 
lightcurves become more sharply peaked at higher redshifts -- 
the sense that might be 
expected if triggering events are more violent and/or 
feedback is more efficient. Evolution in the opposite sense 
(evolution towards the predictions of the stellar wind fueling 
or isolated accretion disk models discussed in \S~\ref{sec:compare.models}, 
or towards less efficient feedback) is ruled out. 
Together with the constraints above regarding evolution in the 
duty cycle, $\delta\equiv t_{\rm eff}/t_{H} \sim t_{Q}\times{\rm d}N/{\rm d}t$, 
the constraints on evolution in $t_{Q}$ imply corresponding constraints 
on the triggering rate of independent AGN events, ${\rm d}N/{\rm d}t$.

\section{Implications for the Integrated Growth of Black Holes}
\label{sec:implications}

\begin{figure}
    \centering
    \scaleup
    \plotone{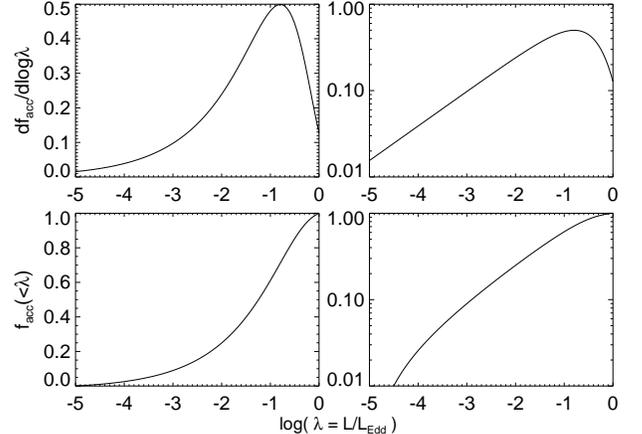}
    \caption{Contribution from different Eddington ratios to the integrated 
    BH growth, given the observationally constrained typical quasar lifetime distribution 
    (Equation~\ref{eqn:schechterfit}). {\em Top:} Differential contribution to the integrated 
    BH mass from each logarithmic interval in Eddington ratio $\lambda$, 
    shown with a linear ({\em left}) and logarithmic ({\em right}) y-axis scale. 
    {\em Bottom:} Cumulative contribution from all Eddington ratios less than 
    the given $\lambda$. From the observations, the BH growth is dominated 
    by moderate/large Eddington ratios $\sim0.2$. The contribution from 
    low Eddington ratios that may be radiatively inefficient ($\lambda\lesssim0.01$) 
    is small ($\sim20\%$), but non-negligible, in agreement with 
    various independent constraints \citep[see][]{hopkins:old.age}. 
    \label{fig:growth}}
\end{figure}

Given the observationally constrained AGN lifetime distribution, Figure~\ref{fig:growth}  
uses it to infer how different Eddington ratios contribute to integrated 
black hole growth. Essentially this amounts to assuming that the shape of 
the Eddington ratio distribution is well-described by Equation~\ref{eqn:schechterfit}, 
and integrating over the time at each Eddington ratio to determine the 
fractional contribution to the final BH mass from each range in $\mdot$. 
Recall from \S~\ref{sec:compare.models}, the total mass accreted must add to 
the total BH mass, and so in fractional terms the absolute 
normalization of the lifetimes can be factored out. Also as noted above, 
it makes little difference (because little mass growth is contributed by such low accretion 
rates) whether we truncate the distribution at some sufficiently low $\mdot$ or 
integrate down to $\mdot\rightarrow0$. For simplicity, we consider the median 
best-fit lifetime distribution with $\beta\approx0.6$, the dependence on mass 
that may be present is sufficiently weak that the results are similar in the observed range. 

Figure~\ref{fig:growth} shows the fractional contribution to the final BH mass 
from each logarithmic interval in $\mdot$ (as well as the cumulative 
contribution from all Eddington ratios $<\mdot$). Given the observations, 
BH growth is dominated 
by moderate/large Eddington ratios $\sim0.2$. 
The contribution from 
low Eddington ratios that may be radiatively inefficient ($\lambda\lesssim0.01$) 
is small ($\sim20\%$), but it is worth noting that it is 
not entirely negligible, and this fraction is sufficient that such populations could 
be a significant contributor to the growth in some low-luminosity AGN populations 
\citep[for more detailed comparison, we refer to][]{hopkins:seyfert.bimodality}. 
These expectations are in good agreement with various independent 
constraints, integral arguments \citep{soltan82}, 
and models for the fueling of AGN and buildup of the 
BH mass function \citep[for a review of these constraints and discussion of 
the contribution of radiatively inefficient sources, see][and references therein]{hopkins:old.age}. 

For this reason, it is easy to construct models matching both the observed 
QLF and the $z=0$ BH mass function with a ``universal'' lightcurve within the 
constraints developed here. 
For the specific lightcurve models discussed here, 
\citet{hopkins:bol.qlf} show this explicitly (see their Figure~10, 
comparing the predicted BH mass function from fitting the compiled QLF 
data with the same set of light curves as those considered here). 
Similar conclusions are reached by \citet{yulu:lightcurve.constraints.from.bhmf.integration}.

\section{Discussion and Conclusions}
\label{sec:discussion}

\subsection{Overview}
\label{sec:discussion:overview}

We have compared observations of the Eddington ratio distribution 
as a function of BH mass, redshift, 
and luminosity with various theories. 
We find good agreement between the observations and the 
predictions of the self-regulated feedback models for BH growth and 
evolution from \citet{hopkins:lifetimes.letter,hopkins:lifetimes.methods,
hopkins:lifetimes.interp,hopkins:qso.all,hopkins:faint.slope,hopkins:seyferts}. 
The agreement covers the entire observed dynamic range, with 
observations extending down $5-6$ orders of 
magnitude in Eddington ratio ($L/L_{\rm Edd} \sim 10^{-6}-1$), 
over three orders of magnitude in BH mass 
($M_{\rm BH}\sim10^{6}-10^{9}\,\msun$), corresponding to 
$\sim8$ orders of magnitude in luminosity, and 
(given indirect observational constraints) from redshifts $z\sim0-1$. 
The models are not fitted to these observations -- there are no free parameters 
adjustable, and the range of uncertainty between different versions of the 
models constrained to reproduce the observed quasar luminosity function 
is small (within the observational error bars). 

The lifetime -- already as constrained in a purely empirical 
sense from the observations, is clearly {\em not} a delta function -- i.e.\ quasars are 
not ``light bulbs'' -- rather, it is a smooth, continuous, 
and relatively steep function of luminosity/Eddington 
ratio, with more time spent at lower luminosities/Eddington ratios. 

We also show
agreement with observed Eddington ratio distributions as a function of AGN 
luminosity, but these 
are less constraining,
because of the inherent Eddington ratio limits 
implied by the selection \citep[wavelength-dependent effects further 
complicate comparisons; discussed in][]{hopkins:seyfert.bimodality}.

\subsection{Implied AGN Lifetimes and Growth Histories}
\label{sec:discussion:constraints}

The observationally implied lifetime distribution can be generally parameterized 
(in model-independent fashion) as a Schechter function, 
with a characteristic normalization lifetime, a turnover at Eddington ratios 
near $\sim1$ (reflecting a physical limit around the Eddington limit), 
and a faint-end slope $\beta$ (such that the luminosity-dependent 
lifetime $t(>\mdot)\propto \mdot^{-\beta}$ at small $\mdot$). 
The observations favor a best-fit $\beta\approx0.6\pm0.05$ (in agreement 
with feedback-regulated models, discussed below) for typical 
$\sim L_{\ast}$ galaxies and BHs. Combined with constraints from the 
quasar luminosity function, there is evidence for weak BH mass-dependence 
of $\beta$, as predicted by hydrodynamic simulations \citep{hopkins:faint.slope}. 

The observations directly yield quasar duty cycles and lifetimes as a 
function of Eddington ratio and BH mass. In terms of the 
integrated time at a given Eddington ratio interval (around some final BH mass), 
the ``quasar'' lifetime -- i.e.\ lifetime at high Eddington ratios $\gtrsim0.1$, is similar 
to that expected from the \citet{salpeter64} time and other observational 
constraints, $\sim10^{8}\,$yr. However, the time at lower Eddington ratios rises 
rapidly -- with $t(\mdot\gtrsim0.01)\sim 0.5-1\,$\,Gyr 
and $t(\mdot \gtrsim 0.001)\sim1-5\,$Gyr.

This allows us to quantify the fractional contribution to 
present-day BH mass from various intervals in $\mdot$; growth is 
dominated by large $\mdot\sim0.2$, with a small -- but non-negligible -- contribution 
$\sim20\%$ from accretion at low Eddington ratios $\mdot\lesssim0.01$, 
in agreement with integral arguments \citep{soltan82} and various independent 
constraints \citep[][and references therein]{hopkins:old.age}.

Comparison of these integral constraints with the $z=0$ observations 
(comparing ``integrated'' and ``effective'' AGN lifetimes/duty cycles) reflects 
the increasingly established AGN downsizing trend: the low-mass BH population 
is still growing today, the high-mass population has shut down since its earlier 
epoch of peak activity.

\subsection{Constraints on Models of Lightcurves and Lifetimes}
\label{sec:discussion:models}

There is enormous constraining power in the 
Eddington ratio distribution at low $\mdot$ in BH mass-limited samples. 
Already, the observations at $z=0$ are sufficient to limit the 
Eddington ratio/quasar lifetime distribution to a narrow range around the 
theoretical predictions from recent hydrodynamic simulations incorporating a 
self-consistent model for accretion and feedback. 
Those models predict that the self-regulated 
nature of BH accretion should lead to a relatively self-similar decay phase in 
AGN luminosity or Eddington ratio (regardless of e.g.\ triggering mechanisms, 
or the exact details of feedback physics), $L\propto t^{-(1.5-2.0)}$, 
which gives rise to the observed power-law-like faint behavior 
in the lifetime distribution with predicted slope $\beta\approx0.6$. 

The observations strongly rule out (at $>5\,\sigma$ significance) 
simplified models for quasar lightcurves, including: ``light-bulb'' models in which quasars 
turn ``on'' for some time at a fixed or relatively narrow range of 
accretion rates, or ``exponential'' models in which AGN 
grow at fixed Eddington ratio and then ``shut off'' rapidly. 

Models where BH accretion simply traces stellar mass loss are also ruled 
out ($\gtrsim4\,\sigma$). Stellar mass loss may still be a fuel source; however, 
the observations argue that some process must further regulate AGN activity, 
shutting down accretion more efficiently than the simple slow starvation expected 
if BH growth directly traced stellar mass loss. 

At $\sim3-5\,\sigma$ significance, 
the observations also rule out isolated accretion disk models 
\citep[][and references therein]{yu:mdot.dist} - i.e.\ accretion disks 
fueled rapidly but then cut off from a future gas supply, without 
feedback (such that they evolve by gas exhaustion). These are a 
considerable improvement on the light bulb model, but are still 
ruled out formally -- with the sense again that some process must 
shut down growth more efficiently (possibly the addition of even 
mild outflows to the isolated disk solution) -- especially for 
high-mass BH populations. 

Local observations of the Eddington ratio distribution alone do not strongly 
constrain the mass or redshift dependence of these properties. Combining 
the observations here with constraints from the quasar luminosity function, 
observations do favor a weak mass-dependence in the shape of the 
lifetime distribution, with the sense that $\beta$ decreases with 
increasing BH mass ($\beta\sim0.6 - 0.2\{\log{[M_{\rm BH}/10^{7}\,\msun]}\}$). 
This is equivalent to the statement that more massive 
systems ``shut off'' more abruptly than low-mass systems; a trend 
predicted by hydrodynamic models, as quasar feedback 
becomes relatively more dominant (and bulge-to-disk ratios increase 
while global disk gas fractions decrease) in more massive systems. 

Recently, \citet{kauffmann:new.mdot.dist} have shown that the $\lambda$ distribution 
depends on the stellar population age of the system. 
We show that the observed trends naturally arise from a {\em single} lightcurve 
of the form constrained here, as a consequence of different triggering histories. 
Systems which are still active/growing (i.e.\ have not ``quenched'' their gas supply, 
BH growth, or star formation) remain at higher accretion rates reflecting the 
median lightcurve directly, while systems which quenched at some earlier 
epoch have decayed down to the lower $\lambda$ power-law-like tail of the 
lightcurve distribution. The existence of these trends, and observational 
constraints on the Eddington ratio distribution at $z=1$ 
\citep{merloni:mdot.dist.fits.prep}, set some constraints on the evolution of 
typical lightcurves with redshift. 

We find that $\beta$ 
and the episodic quasar lifetime $t_{Q}$ cannot evolve strongly 
(parameterized as $\beta\propto(1+z)^{\beta^{\prime}_{z}}$ 
and $t_{Q}\propto(1+z)^{t^{\prime}_{z}}$; the allowed range is 
$-0.5\le \beta^{\prime}_{z} \le 0.05$, $-0.7\le t^{\prime}_{z} \le 0.25$). 
Evolution towards a stellar wind or isolated accretion disk solution is 
strongly ruled out -- rather, the sense of (mild) evolution allowed is 
towards more sharply peaked lightcurves at higher redshift, which 
may be expected if 
fueling is more violent and/or feedback is more efficient in this regime. 
The evolution of duty cycles (and correspondingly 
``effective'' AGN lifetimes) at the high-luminosity end ($\lambda\gtrsim0.1$) 
is directly constrained 
by evolution of the quasar luminosity function -- this limits the 
combination of $t_{0}$ and $\eta$ (but not the faint-end slope $\beta$), 
and (given some constraint on the evolution of episodic lifetimes $t_{Q}$) 
the triggering rate of quasars to evolve according to 
Equations~\ref{eqn:z.evol}-\ref{eqn:z.evol.alpha}, increasing with redshift 
more rapidly in higher-mass BHs. 

These constraints on the AGN can be transformed into 
constraints on the average AGN lightcurve, suggesting a characteristic 
power-law like decay ($L\propto t^{-(1.5-2.0)}$) similar to that predicted 
in models, after rapid growth to some peak luminosity. The observations 
of the Eddington ratio distribution tightly limit the shape of 
such lightcurves, but only weakly constrain the characteristic 
timescale for a single such ``event'' (i.e.\ the width in time of a single ``peak'') -- 
the episodic AGN lifetime -- by setting an upper limit. 

Observational constraints from e.g.\ the 
proximity effect in bright quasars can independently set lower limits on the 
episodic lifetime, however, and suggest that (at least over the observed 
range for the proximity effect, namely quasar activity with 
$\mdot\sim0.1-1$) the episodic lifetime is comparable to the integrated 
lifetime, in the range $t_{Q}\sim0.3-1.0\,t_{\rm int}$. This implies that {\em the typical 
massive BH has experienced no more 
than $\sim$ a couple of bright, high Eddington ratio quasar episodes while 
near its current ($z=0$) mass}. This is in agreement with predictions from 
cosmological models that associate the brightest quasar activity with fueling in violent 
major mergers \citep{hopkins:groups.qso}, for which there are only a couple of 
events expected since high redshift. 
Improved constraints on the episodic lifetime will allow the observed quasar 
luminosity function to be more robustly translated into the distribution of 
AGN triggering rates as a function of BH and host mass. 
Extending the observational constraints on the episodic lifetime 
to lower luminosities/Eddington ratios is also important, as one might expect 
that although a typical object only experiences a couple of triggers to 
near-peak activity, it could have many more triggers for lower-level activity.

\subsection{Other Tests and Future Work}
\label{sec:discussion:future}

Our findings agree with other (less direct) independent 
constraints. Recently, for 
example, \citet{yulu:lightcurve.constraints.from.bhmf.integration} showed that 
the joint evolution of AGN luminosity functions with redshift
favors similar lightcurves. These results (from the QLF and/or BH mass
function and integral/continuity arguments) provide independent support 
for the constraints here, but are primarily sensitive to luminous 
(high-$\lambda$) behavior and cannot distinguish between similar 
lightcurves (with slightly different $\beta$), such as the isolated 
accretion disk or feedback-regulated predictions. The lightcurve shape 
can also be probed by the dependence of AGN clustering on luminosity 
and shape of the QLF or ``active'' BH mass function in deep samples; 
the observations at present appear to favor 
feedback-regulated models over simplified 
light-bulb or exponential lightcurve models \citep[see][]{adelbergersteidel:lifetimes,
myers:clustering,greene:active.mf,daangela:clustering}, but again deeper 
observations are needed to distinguish between 
the proposed physically motivated models. 

Extending the observations of the Eddington ratio distribution to higher redshift 
will greatly improve these constraints as well as limit possible 
redshift evolution in quasar light curves. More massive black holes will be closer to their 
peak growth at higher redshifts, allowing observations to probe the times 
of greatest interest. At $z\sim1-3$, the $\mdot\gtrsim0.1$ end of the 
distribution, in broad-line luminous quasars, is already constrained by the 
observed QLF and application of the virial BH mass estimators 
\citep{kollmeier:mdot,fine:broadline.distrib}. 
Combined with these constraints, smaller volume but deep redshift surveys 
can be used to construct samples which are complete to a given BH/bulge mass, 
and similar narrow-line searches in these hosts could limit the Eddington ratio 
distribution at much lower ratios. If 
coverage is sufficient, X-ray data can be used in the same manner. 
At $z\ge2$, complete, large-volume spectroscopic or X-ray samples are not available 
at present. Here, indirect tests will remain important for the near future, 
but improved constraints on 
what (if any) evolution is seen at lower 
redshifts in the shape of the lifetime distributions 
will considerably inform future models and observational efforts.

\acknowledgments We thank 
Josh Younger, T.~J.\ Cox, and Guinevere Kauffmann 
for helpful discussions. This work
was supported in part by NSF grants ACI 96-19019, AST 00-71019, AST 02-06299, 
and AST 03-07690, and NASA ATP grants NAG5-12140, NAG5-13292, and NAG5-13381. 
Support for PFH was provided by the Miller Institute for Basic Research 
in Science, University of California Berkeley.

\bibliography{/Users/phopkins/Documents/lars_galaxies/papers/ms}

\begin{thebibliography}{162}
\expandafter\ifx\csname natexlab\endcsname\relax\def\natexlab#1{#1}\fi

\bibitem[{{Adelberger} \& {Steidel}(2005)}]{adelbergersteidel:lifetimes}
{Adelberger}, K.~L., \& {Steidel}, C.~C. 2005, \apj, 630, 50

\bibitem[{{Aller} \& {Richstone}(2007)}]{aller:mbh.esph}
{Aller}, M.~C., \& {Richstone}, D.~O. 2007, \apj, 665, 120

\bibitem[{{Bahcall} {et~al.}(1997){Bahcall}, {Kirhakos}, {Saxe}, \&
  {Schneider}}]{bahcall:qso.hosts}
{Bahcall}, J.~N., {Kirhakos}, S., {Saxe}, D.~H., \& {Schneider}, D.~P. 1997,
  \apj, 479, 642

\bibitem[{{Bajtlik} {et~al.}(1988){Bajtlik}, {Duncan}, \&
  {Ostriker}}]{bajtlik:gunnpetersen.qso.age.est}
{Bajtlik}, S., {Duncan}, R.~C., \& {Ostriker}, J.~P. 1988, \apj, 327, 570

\bibitem[{{Barger} \& {Cowie}(2005)}]{barger:qlf}
{Barger}, A.~J., \& {Cowie}, L.~L. 2005, \apj, 635, 115

\bibitem[{{Bassani} {et~al.}(2006)}]{bassani:integral.qlf}
{Bassani}, L., {et~al.} 2006, \apjl, 636, L65

\bibitem[{{Beckmann} {et~al.}(2006){Beckmann}, {Soldi}, {Shrader}, {Gehrels},
  \& {Produit}}]{beckmann:very.hx.qlf}
{Beckmann}, V., {Soldi}, S., {Shrader}, C.~R., {Gehrels}, N., \& {Produit}, N.
  2006, \apj, 652, 126

\bibitem[{{Bell} \& {de Jong}(2000)}]{belldejong:disk.sfh}
{Bell}, E.~F., \& {de Jong}, R.~S. 2000, \mnras, 312, 497

\bibitem[{{Bell} {et~al.}(2003){Bell}, {McIntosh}, {Katz}, \&
  {Weinberg}}]{bell:mfs}
{Bell}, E.~F., {McIntosh}, D.~H., {Katz}, N., \& {Weinberg}, M.~D. 2003, \apjs,
  149, 289

\bibitem[{{Bennert} {et~al.}(2002){Bennert}, {Falcke}, {Schulz}, {Wilson}, \&
  {Wills}}]{bennert:nlr.structure}
{Bennert}, N., {Falcke}, H., {Schulz}, H., {Wilson}, A.~S., \& {Wills}, B.~J.
  2002, \apjl, 574, L105

\bibitem[{{Bird} {et~al.}(2008){Bird}, {Martini}, \&
  {Kaiser}}]{bird:bright.jet.lifetimes}
{Bird}, J., {Martini}, P., \& {Kaiser}, C. 2008, \apj, 676, 147

\bibitem[{{Blundell} {et~al.}(1999){Blundell}, {Rawlings}, \&
  {Willott}}]{blundell:radio.jet.size.lifetimes}
{Blundell}, K.~M., {Rawlings}, S., \& {Willott}, C.~J. 1999, \aj, 117, 677

\bibitem[{{Bruzual} \& {Charlot}(2003)}]{BC03}
{Bruzual}, G., \& {Charlot}, S. 2003, \mnras, 344, 1000

\bibitem[{{Bundy} {et~al.}(2006)}]{bundy:mtrans}
{Bundy}, K., {et~al.} 2006, \apj, 651, 120

\bibitem[{{Bundy} {et~al.}(2008)}]{bundy:agn.lf.to.mf.evol}
---. 2008, \apj, 681, 931

\bibitem[{{Burkert} \& {Silk}(2001)}]{burkertsilk:msigma}
{Burkert}, A., \& {Silk}, J. 2001, \apjl, 554, L151

\bibitem[{{Canalizo} \&
  {Stockton}(2001)}]{canalizostockton01:postsb.qso.mergers}
{Canalizo}, G., \& {Stockton}, A. 2001, \apj, 555, 719

\bibitem[{{Cao} \& {Xu}(2007)}]{cao:riaf.constraints}
{Cao}, X., \& {Xu}, Y.-D. 2007, \mnras, 377, 425

\bibitem[{{Ciotti} \& {Ostriker}(1997)}]{ciottiostriker:cooling.flow.selfreg.1}
{Ciotti}, L., \& {Ostriker}, J.~P. 1997, \apjl, 487, L105+

\bibitem[{{Ciotti} \& {Ostriker}(2001)}]{ciottiostriker:cooling.flow.selfreg.2}
---. 2001, \apj, 551, 131

\bibitem[{{Ciotti} \& {Ostriker}(2007)}]{ciottiostriker:recycling}
---. 2007, \apj, 665, 1038

\bibitem[{{Coil} {et~al.}(2007){Coil}, {Hennawi}, {Newman}, {Cooper}, \&
  {Davis}}]{coil:agn.clustering}
{Coil}, A.~L., {Hennawi}, J.~F., {Newman}, J.~A., {Cooper}, M.~C., \& {Davis},
  M. 2007, \apj, 654, 115

\bibitem[{{Croom} {et~al.}(2005){Croom}, {Boyle}, {Shanks}, {Smith}, {Miller},
  {Outram}, {Loaring}, {Hoyle}, \& {da {\^A}ngela}}]{croom:clustering}
{Croom}, S.~M., {Boyle}, B.~J., {Shanks}, T., {Smith}, R.~J., {Miller}, L.,
  {Outram}, P.~J., {Loaring}, N.~S., {Hoyle}, F., \& {da {\^A}ngela}, J. 2005,
  \mnras, 356, 415

\bibitem[{{da Angela} {et~al.}(2008)}]{daangela:clustering}
{da Angela}, J., {et~al.} 2008, \mnras, 383, 565

\bibitem[{{Di Matteo} {et~al.}(2008){Di Matteo}, {Colberg}, {Springel},
  {Hernquist}, \& {Sijacki}}]{dimatteo:cosmo.bhs}
{Di Matteo}, T., {Colberg}, J., {Springel}, V., {Hernquist}, L., \& {Sijacki},
  D. 2008, \apj, 676, 33

\bibitem[{{Di Matteo} {et~al.}(2003){Di Matteo}, {Croft}, {Springel}, \&
  {Hernquist}}]{dimatteo:cosmo.bh.growth.sim.1}
{Di Matteo}, T., {Croft}, R.~A.~C., {Springel}, V., \& {Hernquist}, L. 2003,
  \apj, 593, 56

\bibitem[{{Di Matteo} {et~al.}(2004){Di Matteo}, {Croft}, {Springel}, \&
  {Hernquist}}]{dimatteo:qso.host.metals}
---. 2004, \apj, 610, 80

\bibitem[{{Di Matteo} {et~al.}(2005){Di Matteo}, {Springel}, \&
  {Hernquist}}]{dimatteo:msigma}
{Di Matteo}, T., {Springel}, V., \& {Hernquist}, L. 2005, \nat, 433, 604

\bibitem[{{Dunlop} {et~al.}(2003){Dunlop}, {McLure}, {Kukula}, {Baum}, {O'Dea},
  \& {Hughes}}]{dunlop:qso.hosts}
{Dunlop}, J.~S., {McLure}, R.~J., {Kukula}, M.~J., {Baum}, S.~A., {O'Dea},
  C.~P., \& {Hughes}, D.~H. 2003, \mnras, 340, 1095

\bibitem[{{Elvis} {et~al.}(1994)}]{elvis:atlas}
{Elvis}, M., {et~al.} 1994, \apjs, 95, 1

\bibitem[{{Erb} {et~al.}(2006){Erb}, {Steidel}, {Shapley}, {Pettini}, {Reddy},
  \& {Adelberger}}]{erb:lbg.gasmasses}
{Erb}, D.~K., {Steidel}, C.~C., {Shapley}, A.~E., {Pettini}, M., {Reddy},
  N.~A., \& {Adelberger}, K.~L. 2006, \apj, 646, 107

\bibitem[{{Falcke} \& {Biermann}(1996)}]{falcke:radio.vs.mdot}
{Falcke}, H., \& {Biermann}, P.~L. 1996, \aap, 308, 321

\bibitem[{{Falcke} {et~al.}(2004){Falcke}, {K{\"o}rding}, \&
  {Markoff}}]{falcke04:radio.vs.mdot}
{Falcke}, H., {K{\"o}rding}, E., \& {Markoff}, S. 2004, \aap, 414, 895

\bibitem[{{Faucher-Gigu{\`e}re}
  {et~al.}(2008{\natexlab{a}}){Faucher-Gigu{\`e}re}, {Lidz}, {Hernquist}, \&
  {Zaldarriaga}}]{faucher:heII.reion.qso.constraints}
{Faucher-Gigu{\`e}re}, C.-A., {Lidz}, A., {Hernquist}, L., \& {Zaldarriaga}, M.
  2008{\natexlab{a}}, \apjl, 682, L9

\bibitem[{{Faucher-Gigu{\`e}re}
  {et~al.}(2008{\natexlab{b}}){Faucher-Gigu{\`e}re}, {Lidz}, {Hernquist}, \&
  {Zaldarriaga}}]{faucher:ion.background.evol}
---. 2008{\natexlab{b}}, \apj, 688, 85

\bibitem[{{Faucher-Gigu{\`e}re}
  {et~al.}(2008{\natexlab{c}}){Faucher-Gigu{\`e}re}, {Lidz}, {Zaldarriaga}, \&
  {Hernquist}}]{faucher:proximity}
{Faucher-Gigu{\`e}re}, C.-A., {Lidz}, A., {Zaldarriaga}, M., \& {Hernquist}, L.
  2008{\natexlab{c}}, \apj, 673, 39

\bibitem[{{Fender} {et~al.}(2007){Fender}, {Koerding}, {Belloni}, {Uttley},
  {McHardy}, \& {Tzioumis}}]{fender:radio.mdot.review}
{Fender}, R., {Koerding}, E., {Belloni}, T., {Uttley}, P., {McHardy}, I., \&
  {Tzioumis}, T. 2007, ArXiv e-prints, in press, arXiv:0706.3838, 706

\bibitem[{{Ferrarese} \& {Merritt}(2000)}]{FM00}
{Ferrarese}, L., \& {Merritt}, D. 2000, \apjl, 539, L9

\bibitem[{{Fine} {et~al.}(2006)}]{fine:mbh-mhalo.clustering}
{Fine}, S., {et~al.} 2006, \mnras, 373, 613

\bibitem[{{Fine} {et~al.}(2008)}]{fine:broadline.distrib}
---. 2008, \mnras, 390, 1413

\bibitem[{{Fiore} {et~al.}(2003)}]{fiore:type2.lx.vs.lhost}
{Fiore}, F., {et~al.} 2003, \aap, 409, 79

\bibitem[{{Fontanot} {et~al.}(2007){Fontanot}, {Cristiani}, {Monaco}, {Nonino},
  {Vanzella}, {Brandt}, {Grazian}, \& {Mao}}]{fontanot:highz.qlf}
{Fontanot}, F., {Cristiani}, S., {Monaco}, P., {Nonino}, M., {Vanzella}, E.,
  {Brandt}, W.~N., {Grazian}, A., \& {Mao}, J. 2007, \aap, 461, 39

\bibitem[{{Gebhardt} {et~al.}(2000)}]{Gebhardt00}
{Gebhardt}, K., {et~al.} 2000, \apjl, 539, L13

\bibitem[{{Gilli} {et~al.}(2007){Gilli}, {Comastri}, \&
  {Hasinger}}]{gilli:obscured.fractions}
{Gilli}, R., {Comastri}, A., \& {Hasinger}, G. 2007, \aap, 463, 79

\bibitem[{{Gon{\c c}alves} {et~al.}(2008){Gon{\c c}alves}, {Steidel}, \&
  {Pettini}}]{goncalves:transverse.proximity}
{Gon{\c c}alves}, T.~S., {Steidel}, C.~C., \& {Pettini}, M. 2008, \apj, 676,
  816

\bibitem[{{Granato} {et~al.}(2004){Granato}, {De Zotti}, {Silva}, {Bressan}, \&
  {Danese}}]{granato:sam}
{Granato}, G.~L., {De Zotti}, G., {Silva}, L., {Bressan}, A., \& {Danese}, L.
  2004, \apj, 600, 580

\bibitem[{{Grazian} {et~al.}(2004){Grazian}, {Negrello}, {Moscardini},
  {Cristiani}, {Haehnelt}, {Matarrese}, {Omizzolo}, \&
  {Vanzella}}]{grazian:local.qso.clustering}
{Grazian}, A., {Negrello}, M., {Moscardini}, L., {Cristiani}, S., {Haehnelt},
  M.~G., {Matarrese}, S., {Omizzolo}, A., \& {Vanzella}, E. 2004, \aj, 127, 592

\bibitem[{{Greene} \& {Ho}(2007)}]{greene:active.mf}
{Greene}, J.~E., \& {Ho}, L.~C. 2007, \apj, 667, 131

\bibitem[{{Greene} {et~al.}(2006){Greene}, {Ho}, \&
  {Ulvestad}}]{greene:radio.vs.edd}
{Greene}, J.~E., {Ho}, L.~C., \& {Ulvestad}, J.~S. 2006, \apj, 636, 56

\bibitem[{{Haehnelt} {et~al.}(1998){Haehnelt}, {Natarajan}, \&
  {Rees}}]{haehnelt:bh.synthesis.model}
{Haehnelt}, M.~G., {Natarajan}, P., \& {Rees}, M.~J. 1998, \mnras, 300, 817

\bibitem[{{Haiman} \& {Cen}(2002)}]{haiman:gunnpetersen.qso.age.est}
{Haiman}, Z., \& {Cen}, R. 2002, \apj, 578, 702

\bibitem[{{Haiman} {et~al.}(2004){Haiman}, {Ciotti}, \&
  {Ostriker}}]{haiman:bhmf}
{Haiman}, Z., {Ciotti}, L., \& {Ostriker}, J.~P. 2004, \apj, 606, 763

\bibitem[{{Haiman} {et~al.}(2007){Haiman}, {Jimenez}, \&
  {Bernardi}}]{haiman:qlf.from.ell.ages}
{Haiman}, Z., {Jimenez}, R., \& {Bernardi}, M. 2007, \apj, 658, 721

\bibitem[{{Hamilton} {et~al.}(2002){Hamilton}, {Casertano}, \&
  {Turnshek}}]{hamilton:qso.host.lf}
{Hamilton}, T.~S., {Casertano}, S., \& {Turnshek}, D.~A. 2002, \apj, 576, 61

\bibitem[{{Hasinger}(2008)}]{hasinger:absorption.update}
{Hasinger}, G. 2008, \aap, 490, 905

\bibitem[{{Hasinger} {et~al.}(2005){Hasinger}, {Miyaji}, \&
  {Schmidt}}]{hasinger05:qlf}
{Hasinger}, G., {Miyaji}, T., \& {Schmidt}, M. 2005, \aap, 441, 417

\bibitem[{{Heckman} {et~al.}(2004){Heckman}, {Kauffmann}, {Brinchmann},
  {Charlot}, {Tremonti}, \& {White}}]{heckman:local.mbh}
{Heckman}, T.~M., {Kauffmann}, G., {Brinchmann}, J., {Charlot}, S., {Tremonti},
  C., \& {White}, S.~D.~M. 2004, \apj, 613, 109

\bibitem[{{Hickox} {et~al.}(2007)}]{hickox:bootes.obscured.agn}
{Hickox}, R.~C., {et~al.} 2007, \apj, 671, 1365

\bibitem[{{Hickox} {et~al.}(2009)}]{hickox:multiwavelength.agn}
---. 2009, \apj, 696, 891

\bibitem[{{Ho}(2002)}]{ho:radio.vs.mdot}
{Ho}, L.~C. 2002, \apj, 564, 120

\bibitem[{{Hopkins} {et~al.}(2007{\natexlab{a}}){Hopkins}, {Bundy},
  {Hernquist}, \& {Ellis}}]{hopkins:transition.mass}
{Hopkins}, P.~F., {Bundy}, K., {Hernquist}, L., \& {Ellis}, R.~S.
  2007{\natexlab{a}}, \apj, 659, 976

\bibitem[{{Hopkins} {et~al.}(2009{\natexlab{a}}){Hopkins}, {Cox}, {Dutta},
  {Hernquist}, {Kormendy}, \& {Lauer}}]{hopkins:cusps.ell}
{Hopkins}, P.~F., {Cox}, T.~J., {Dutta}, S.~N., {Hernquist}, L., {Kormendy},
  J., \& {Lauer}, T.~R. 2009{\natexlab{a}}, \apjs, 181, 135

\bibitem[{{Hopkins} {et~al.}(2008{\natexlab{a}}){Hopkins}, {Cox}, \&
  {Hernquist}}]{hopkins:cusps.fp}
{Hopkins}, P.~F., {Cox}, T.~J., \& {Hernquist}, L. 2008{\natexlab{a}}, \apj,
  689, 17

\bibitem[{{Hopkins} {et~al.}(2008{\natexlab{b}}){Hopkins}, {Cox}, {Kere{\v s}},
  \& {Hernquist}}]{hopkins:groups.ell}
{Hopkins}, P.~F., {Cox}, T.~J., {Kere{\v s}}, D., \& {Hernquist}, L.
  2008{\natexlab{b}}, \apjs, 175, 390

\bibitem[{{Hopkins} {et~al.}(2009{\natexlab{b}}){Hopkins}, {Cox}, {Younger}, \&
  {Hernquist}}]{hopkins:disk.survival}
{Hopkins}, P.~F., {Cox}, T.~J., {Younger}, J.~D., \& {Hernquist}, L.
  2009{\natexlab{b}}, \apj, 691, 1168

\bibitem[{{Hopkins} \& {Hernquist}(2006)}]{hopkins:seyferts}
{Hopkins}, P.~F., \& {Hernquist}, L. 2006, \apjs, 166, 1

\bibitem[{{Hopkins} \& {Hernquist}(2009)}]{hopkins:seyfert.limits}
---. 2009, \apj, 694, 599

\bibitem[{{Hopkins} {et~al.}(2005{\natexlab{a}}){Hopkins}, {Hernquist}, {Cox},
  {Di Matteo}, {Martini}, {Robertson}, \&
  {Springel}}]{hopkins:lifetimes.methods}
{Hopkins}, P.~F., {Hernquist}, L., {Cox}, T.~J., {Di Matteo}, T., {Martini},
  P., {Robertson}, B., \& {Springel}, V. 2005{\natexlab{a}}, \apj, 630, 705

\bibitem[{{Hopkins} {et~al.}(2005{\natexlab{b}}){Hopkins}, {Hernquist}, {Cox},
  {Di Matteo}, {Robertson}, \& {Springel}}]{hopkins:lifetimes.interp}
{Hopkins}, P.~F., {Hernquist}, L., {Cox}, T.~J., {Di Matteo}, T., {Robertson},
  B., \& {Springel}, V. 2005{\natexlab{b}}, \apj, 630, 716

\bibitem[{{Hopkins} {et~al.}(2005{\natexlab{c}}){Hopkins}, {Hernquist}, {Cox},
  {Di Matteo}, {Robertson}, \& {Springel}}]{hopkins:lifetimes.obscuration}
---. 2005{\natexlab{c}}, \apj, 632, 81

\bibitem[{{Hopkins} {et~al.}(2006{\natexlab{a}}){Hopkins}, {Hernquist}, {Cox},
  {Di Matteo}, {Robertson}, \& {Springel}}]{hopkins:qso.all}
---. 2006{\natexlab{a}}, \apjs, 163, 1

\bibitem[{{Hopkins} {et~al.}(2008{\natexlab{c}}){Hopkins}, {Hernquist}, {Cox},
  {Dutta}, \& {Rothberg}}]{hopkins:cusps.mergers}
{Hopkins}, P.~F., {Hernquist}, L., {Cox}, T.~J., {Dutta}, S.~N., \& {Rothberg},
  B. 2008{\natexlab{c}}, \apj, 679, 156

\bibitem[{{Hopkins} {et~al.}(2008{\natexlab{d}}){Hopkins}, {Hernquist}, {Cox},
  \& {Kere{\v s}}}]{hopkins:groups.qso}
{Hopkins}, P.~F., {Hernquist}, L., {Cox}, T.~J., \& {Kere{\v s}}, D.
  2008{\natexlab{d}}, \apjs, 175, 356

\bibitem[{{Hopkins} {et~al.}(2009{\natexlab{c}}){Hopkins}, {Hernquist}, {Cox},
  {Kere{\v s}}, \& {Wuyts}}]{hopkins:cusps.evol}
{Hopkins}, P.~F., {Hernquist}, L., {Cox}, T.~J., {Kere{\v s}}, D., \& {Wuyts},
  S. 2009{\natexlab{c}}, \apj, 691, 1424

\bibitem[{{Hopkins} {et~al.}(2006{\natexlab{b}}){Hopkins}, {Hernquist}, {Cox},
  {Robertson}, {Di Matteo}, \& {Springel}}]{hopkins:faint.slope}
{Hopkins}, P.~F., {Hernquist}, L., {Cox}, T.~J., {Robertson}, B., {Di Matteo},
  T., \& {Springel}, V. 2006{\natexlab{b}}, \apj, 639, 700

\bibitem[{{Hopkins} {et~al.}(2007{\natexlab{b}}){Hopkins}, {Hernquist}, {Cox},
  {Robertson}, \& {Krause}}]{hopkins:bhfp.theory}
{Hopkins}, P.~F., {Hernquist}, L., {Cox}, T.~J., {Robertson}, B., \& {Krause},
  E. 2007{\natexlab{b}}, \apj, 669, 45

\bibitem[{{Hopkins} {et~al.}(2007{\natexlab{c}}){Hopkins}, {Hernquist}, {Cox},
  {Robertson}, \& {Krause}}]{hopkins:bhfp.obs}
---. 2007{\natexlab{c}}, \apj, 669, 67

\bibitem[{{Hopkins} {et~al.}(2006{\natexlab{c}}){Hopkins}, {Hernquist}, {Cox},
  {Robertson}, \& {Springel}}]{hopkins:red.galaxies}
{Hopkins}, P.~F., {Hernquist}, L., {Cox}, T.~J., {Robertson}, B., \&
  {Springel}, V. 2006{\natexlab{c}}, \apjs, 163, 50

\bibitem[{{Hopkins} {et~al.}(2005{\natexlab{d}}){Hopkins}, {Hernquist},
  {Martini}, {Cox}, {Robertson}, {Di Matteo}, \&
  {Springel}}]{hopkins:lifetimes.letter}
{Hopkins}, P.~F., {Hernquist}, L., {Martini}, P., {Cox}, T.~J., {Robertson},
  B., {Di Matteo}, T., \& {Springel}, V. 2005{\natexlab{d}}, \apjl, 625, L71

\bibitem[{{Hopkins} {et~al.}(2009{\natexlab{d}}){Hopkins}, {Hickox},
  {Quataert}, \& {Hernquist}}]{hopkins:seyfert.bimodality}
{Hopkins}, P.~F., {Hickox}, R., {Quataert}, E., \& {Hernquist}, L.
  2009{\natexlab{d}}, \mnras, in press, arXiv:0901.2936 [astro-ph]

\bibitem[{{Hopkins} {et~al.}(2009{\natexlab{e}}){Hopkins}, {Lauer}, {Cox},
  {Hernquist}, \& {Kormendy}}]{hopkins:cores}
{Hopkins}, P.~F., {Lauer}, T.~R., {Cox}, T.~J., {Hernquist}, L., \& {Kormendy},
  J. 2009{\natexlab{e}}, \apjs, 181, 486

\bibitem[{{Hopkins} {et~al.}(2007{\natexlab{d}}){Hopkins}, {Lidz}, {Hernquist},
  {Coil}, {Myers}, {Cox}, \& {Spergel}}]{hopkins:clustering}
{Hopkins}, P.~F., {Lidz}, A., {Hernquist}, L., {Coil}, A.~L., {Myers}, A.~D.,
  {Cox}, T.~J., \& {Spergel}, D.~N. 2007{\natexlab{d}}, \apj, 662, 110

\bibitem[{{Hopkins} {et~al.}(2006{\natexlab{d}}){Hopkins}, {Narayan}, \&
  {Hernquist}}]{hopkins:old.age}
{Hopkins}, P.~F., {Narayan}, R., \& {Hernquist}, L. 2006{\natexlab{d}}, \apj,
  643, 641

\bibitem[{{Hopkins} {et~al.}(2007{\natexlab{e}}){Hopkins}, {Richards}, \&
  {Hernquist}}]{hopkins:bol.qlf}
{Hopkins}, P.~F., {Richards}, G.~T., \& {Hernquist}, L. 2007{\natexlab{e}},
  \apj, 654, 731

\bibitem[{{Hopkins} {et~al.}(2006{\natexlab{e}}){Hopkins}, {Somerville},
  {Hernquist}, {Cox}, {Robertson}, \& {Li}}]{hopkins:merger.lfs}
{Hopkins}, P.~F., {Somerville}, R.~S., {Hernquist}, L., {Cox}, T.~J.,
  {Robertson}, B., \& {Li}, Y. 2006{\natexlab{e}}, \apj, 652, 864

\bibitem[{{Jakobsen} {et~al.}(2003){Jakobsen}, {Jansen}, {Wagner}, \&
  {Reimers}}]{jakobsen:heII.ion.transverse.proximity}
{Jakobsen}, P., {Jansen}, R.~A., {Wagner}, S., \& {Reimers}, D. 2003, \aap,
  397, 891

\bibitem[{{Jester}(2005)}]{jester:riaf.test}
{Jester}, S. 2005, \apj, 625, 667

\bibitem[{{Johansson} {et~al.}(2009){Johansson}, {Naab}, \&
  {Burkert}}]{johansson:mixed.morph.mbh.sims}
{Johansson}, P.~H., {Naab}, T., \& {Burkert}, A. 2009, \apj, 690, 802

\bibitem[{{Kauffmann} \& {Haehnelt}(2000)}]{kh00}
{Kauffmann}, G., \& {Haehnelt}, M. 2000, \mnras, 311, 576

\bibitem[{{Kauffmann} \& {Heckman}(2008)}]{kauffmann:new.mdot.dist}
{Kauffmann}, G., \& {Heckman}, T.~M. 2008, \mnras, in press, arXiv:0812.1224

\bibitem[{{Kauffmann} {et~al.}(2003)}]{kauffmann:qso.hosts}
{Kauffmann}, G., {et~al.} 2003, \mnras, 346, 1055

\bibitem[{{Kewley} {et~al.}(2006){Kewley}, {Groves}, {Kauffmann}, \&
  {Heckman}}]{kewley:agn.host.sf}
{Kewley}, L.~J., {Groves}, B., {Kauffmann}, G., \& {Heckman}, T. 2006, \mnras,
  372, 961

\bibitem[{{Kollmeier} {et~al.}(2006)}]{kollmeier:mdot}
{Kollmeier}, J.~A., {et~al.} 2006, \apj, 648, 128

\bibitem[{{Komatsu} {et~al.}(2009)}]{komatsu:wmap5}
{Komatsu}, E., {et~al.} 2009, \apjs, 180, 330

\bibitem[{{Kriek} {et~al.}(2007)}]{kriek:qso.frac}
{Kriek}, M., {et~al.} 2007, \apj, 669, 776

\bibitem[{{La Franca} {et~al.}(2005)}]{lafranca:hx.qlf}
{La Franca}, F., {et~al.} 2005, \apj, 635, 864

\bibitem[{{Lapi} {et~al.}(2006){Lapi}, {Shankar}, {Mao}, {Granato}, {Silva},
  {De Zotti}, \& {Danese}}]{lapi:qlf.sam}
{Lapi}, A., {Shankar}, F., {Mao}, J., {Granato}, G.~L., {Silva}, L., {De
  Zotti}, G., \& {Danese}, L. 2006, \apj, 650, 42

\bibitem[{{Leitherer} {et~al.}(1999)}]{starburst99}
{Leitherer}, C., {et~al.} 1999, \apjs, 123, 3

\bibitem[{{Lidz} {et~al.}(2006){Lidz}, {Hopkins}, {Cox}, {Hernquist}, \&
  {Robertson}}]{lidz:clustering}
{Lidz}, A., {Hopkins}, P.~F., {Cox}, T.~J., {Hernquist}, L., \& {Robertson}, B.
  2006, \apj, 641, 41

\bibitem[{{Lidz} {et~al.}(2007){Lidz}, {McQuinn}, {Zaldarriaga}, {Hernquist},
  \& {Dutta}}]{lidz:proximity}
{Lidz}, A., {McQuinn}, M., {Zaldarriaga}, M., {Hernquist}, L., \& {Dutta}, S.
  2007, \apj, 670, 39

\bibitem[{{Maccarone} {et~al.}(2003){Maccarone}, {Gallo}, \&
  {Fender}}]{maccarone:agn.riaf.connection}
{Maccarone}, T.~J., {Gallo}, E., \& {Fender}, R. 2003, \mnras, 345, L19

\bibitem[{{Magorrian} {et~al.}(1998)}]{magorrian}
{Magorrian}, J., {et~al.} 1998, \aj, 115, 2285

\bibitem[{{Marchesini} {et~al.}(2004){Marchesini}, {Celotti}, \&
  {Ferrarese}}]{marchesini:low.mdot.sample}
{Marchesini}, D., {Celotti}, A., \& {Ferrarese}, L. 2004, \mnras, 351, 733

\bibitem[{{Marconi} \& {Hunt}(2003)}]{marconihunt}
{Marconi}, A., \& {Hunt}, L.~K. 2003, \apjl, 589, L21

\bibitem[{{Marconi} {et~al.}(2004){Marconi}, {Risaliti}, {Gilli}, {Hunt},
  {Maiolino}, \& {Salvati}}]{marconi:bhmf}
{Marconi}, A., {Risaliti}, G., {Gilli}, R., {Hunt}, L.~K., {Maiolino}, R., \&
  {Salvati}, M. 2004, \mnras, 351, 169

\bibitem[{{Martini}(2004)}]{martini04}
{Martini}, P. 2004, in Coevolution of Black Holes and Galaxies, ed. L.~C. C. C.
  U.~P. {Ho}, 169

\bibitem[{{McClintock} \& {Remillard}(2006)}]{mcclintock:xrb.review}
{McClintock}, J.~E., \& {Remillard}, R.~A. 2006, {Black hole binaries} (Compact
  stellar X-ray sources), 157--213

\bibitem[{{McQuinn} {et~al.}(2009){McQuinn}, {Lidz}, {Zaldarriaga},
  {Hernquist}, {Hopkins}, {Dutta}, \&
  {Faucher-Gigu{\`e}re}}]{mcquinn:helium.reionization.model}
{McQuinn}, M., {Lidz}, A., {Zaldarriaga}, M., {Hernquist}, L., {Hopkins},
  P.~F., {Dutta}, S., \& {Faucher-Gigu{\`e}re}, C.-A. 2009, \apj, 694, 842

\bibitem[{{Meier}(2001)}]{meier:jets.in.adaf}
{Meier}, D.~L. 2001, \apjl, 548, L9

\bibitem[{{Menci} {et~al.}(2003){Menci}, {Cavaliere}, {Fontana}, {Giallongo},
  {Poli}, \& {Vittorini}}]{menci:sam}
{Menci}, N., {Cavaliere}, A., {Fontana}, A., {Giallongo}, E., {Poli}, F., \&
  {Vittorini}, V. 2003, \apjl, 587, L63

\bibitem[{{Merloni} \& {Heinz}(2007)}]{merloniheinz.bhfp.radio}
{Merloni}, A., \& {Heinz}, S. 2007, \mnras, 381, 589

\bibitem[{{Merloni} \& {Heinz}(2008)}]{merloni:synthesis.model}
---. 2008, \mnras, 388, 1011

\bibitem[{{Merloni} \& {Heinz}(2009)}]{merloni:mdot.dist.fits.prep}
---. 2009, \mnras, in prep

\bibitem[{{Merloni} {et~al.}(2003){Merloni}, {Heinz}, \& {di
  Matteo}}]{merloni:bhfp.radio}
{Merloni}, A., {Heinz}, S., \& {di Matteo}, T. 2003, \mnras, 345, 1057

\bibitem[{{Murray} {et~al.}(2005){Murray}, {Quataert}, \&
  {Thompson}}]{murray:momentum.winds}
{Murray}, N., {Quataert}, E., \& {Thompson}, T.~A. 2005, \apj, 618, 569

\bibitem[{{Myers} {et~al.}(2007){Myers}, {Brunner}, {Nichol}, {Richards},
  {Schneider}, \& {Bahcall}}]{myers:clustering}
{Myers}, A.~D., {Brunner}, R.~J., {Nichol}, R.~C., {Richards}, G.~T.,
  {Schneider}, D.~P., \& {Bahcall}, N.~A. 2007, \apj, 658, 85

\bibitem[{{Myers} {et~al.}(2006)}]{myers:clustering.old}
{Myers}, A.~D., {et~al.} 2006, \apj, 638, 622

\bibitem[{{Narayan} \& {Yi}(1994)}]{NY94}
{Narayan}, R., \& {Yi}, I. 1994, \apjl, 428, L13

\bibitem[{{Narayan} {et~al.}(1995){Narayan}, {Yi}, \& {Mahadevan}}]{nym95}
{Narayan}, R., {Yi}, I., \& {Mahadevan}, R. 1995, \nat, 374, 623

\bibitem[{{Narayan} {et~al.}(1996){Narayan}, {Yi}, \& {Mahadevan}}]{nym96}
---. 1996, \aaps, 120, C287+

\bibitem[{{Norman} \& {Scoville}(1988)}]{norman:stellar.wind.fueling}
{Norman}, C., \& {Scoville}, N. 1988, \apj, 332, 124

\bibitem[{{Porciani} {et~al.}(2004){Porciani}, {Magliocchetti}, \&
  {Norberg}}]{porciani2004}
{Porciani}, C., {Magliocchetti}, M., \& {Norberg}, P. 2004, \mnras, 355, 1010

\bibitem[{{Porciani} \& {Norberg}(2006)}]{porciani:clustering}
{Porciani}, C., \& {Norberg}, P. 2006, \mnras, 371, 1824

\bibitem[{{Reynolds} \&
  {Begelman}(1997)}]{reynolds.begelman:radio.source.sizes}
{Reynolds}, C.~S., \& {Begelman}, M.~C. 1997, \apjl, 487, L135+

\bibitem[{{Richards} {et~al.}(2006)}]{richards:seds}
{Richards}, G.~T., {et~al.} 2006, \apjs, 166, 470

\bibitem[{{Salpeter}(1964)}]{salpeter64}
{Salpeter}, E.~E. 1964, \apj, 140, 796

\bibitem[{{Salucci} {et~al.}(1999){Salucci}, {Szuszkiewicz}, {Monaco}, \&
  {Danese}}]{salucci:bhmf}
{Salucci}, P., {Szuszkiewicz}, E., {Monaco}, P., \& {Danese}, L. 1999, \mnras,
  307, 637

\bibitem[{{Sazonov} {et~al.}(2005){Sazonov}, {Ostriker}, {Ciotti}, \&
  {Sunyaev}}]{sazonov:radiative.feedback}
{Sazonov}, S.~Y., {Ostriker}, J.~P., {Ciotti}, L., \& {Sunyaev}, R.~A. 2005,
  \mnras, 358, 168

\bibitem[{{Scannapieco} \& {Oh}(2004)}]{scannapieco:sam}
{Scannapieco}, E., \& {Oh}, S.~P. 2004, \apj, 608, 62

\bibitem[{{Scheuer}(1995)}]{scheuer:radio.jet.size.lifetimes}
{Scheuer}, P.~A.~G. 1995, \mnras, 277, 331

\bibitem[{{Schirber} {et~al.}(2004){Schirber}, {Miralda-Escud{\'e}}, \&
  {McDonald}}]{schirber:transverse.proximity}
{Schirber}, M., {Miralda-Escud{\'e}}, J., \& {McDonald}, P. 2004, \apj, 610,
  105

\bibitem[{{Shakura} \& {Sunyaev}(1973)}]{shakurasunyaev73}
{Shakura}, N.~I., \& {Sunyaev}, R.~A. 1973, \aap, 24, 337

\bibitem[{{Shankar} {et~al.}(2009{\natexlab{a}}){Shankar}, {Bernardi}, \&
  {Haiman}}]{shankar:implied.msig.from.gal.ages}
{Shankar}, F., {Bernardi}, M., \& {Haiman}, Z. 2009{\natexlab{a}}, \apj, 694,
  867

\bibitem[{{Shankar} {et~al.}(2004){Shankar}, {Salucci}, {Granato}, {De Zotti},
  \& {Danese}}]{shankar:bhmf}
{Shankar}, F., {Salucci}, P., {Granato}, G.~L., {De Zotti}, G., \& {Danese}, L.
  2004, \mnras, 354, 1020

\bibitem[{{Shankar} {et~al.}(2009{\natexlab{b}}){Shankar}, {Weinberg}, \&
  {Miralda-Escud{\'e}}}]{shankar:bol.qlf}
{Shankar}, F., {Weinberg}, D.~H., \& {Miralda-Escud{\'e}}, J.
  2009{\natexlab{b}}, \apj, 690, 20

\bibitem[{{Shen} {et~al.}(2007)}]{shen:clustering}
{Shen}, Y., {et~al.} 2007, \aj, 133, 2222

\bibitem[{{Siana} {et~al.}(2008)}]{siana:z3.swire.qlf}
{Siana}, B., {et~al.} 2008, \apj, 675, 49

\bibitem[{{Sijacki} {et~al.}(2007){Sijacki}, {Springel}, {di Matteo}, \&
  {Hernquist}}]{sijacki:radio}
{Sijacki}, D., {Springel}, V., {di Matteo}, T., \& {Hernquist}, L. 2007,
  \mnras, 380, 877

\bibitem[{{Silk} \& {Rees}(1998)}]{silkrees:msigma}
{Silk}, J., \& {Rees}, M.~J. 1998, \aap, 331, L1

\bibitem[{{Silverman} {et~al.}(2005)}]{silverman:hx.spacedensity.ldde}
{Silverman}, J.~D., {et~al.} 2005, \apj, 624, 630

\bibitem[{{Silverman} {et~al.}(2008{\natexlab{a}})}]{silverman:qso.hosts}
---. 2008{\natexlab{a}}, \apj, 675, 1025

\bibitem[{{Silverman} {et~al.}(2008{\natexlab{b}})}]{silverman:hx.lf}
---. 2008{\natexlab{b}}, \apj, 679, 118

\bibitem[{{Simpson}(2005)}]{simpson:type1.frac}
{Simpson}, C. 2005, \mnras, 360, 565

\bibitem[{{Sokasian} {et~al.}(2002){Sokasian}, {Abel}, \&
  {Hernquist}}]{sokasian:heII.reion.epoch}
{Sokasian}, A., {Abel}, T., \& {Hernquist}, L. 2002, \mnras, 332, 601

\bibitem[{{Sokasian} {et~al.}(2003){Sokasian}, {Abel}, \&
  {Hernquist}}]{sokasian:heII.reion}
---. 2003, \mnras, 340, 473

\bibitem[{{Soltan}(1982)}]{soltan82}
{Soltan}, A. 1982, \mnras, 200, 115

\bibitem[{{Springel} {et~al.}(2005{\natexlab{a}}){Springel}, {Di Matteo}, \&
  {Hernquist}}]{springel:red.galaxies}
{Springel}, V., {Di Matteo}, T., \& {Hernquist}, L. 2005{\natexlab{a}}, \apjl,
  620, L79

\bibitem[{{Springel} {et~al.}(2005{\natexlab{b}}){Springel}, {Di Matteo}, \&
  {Hernquist}}]{springel:models}
---. 2005{\natexlab{b}}, \mnras, 361, 776

\bibitem[{{Ueda} {et~al.}(2003){Ueda}, {Akiyama}, {Ohta}, \&
  {Miyaji}}]{ueda03:qlf}
{Ueda}, Y., {Akiyama}, M., {Ohta}, K., \& {Miyaji}, T. 2003, \apj, 598, 886

\bibitem[{{Vestergaard} \&
  {Peterson}(2006)}]{vestergaardpeterson:virial.corr.review}
{Vestergaard}, M., \& {Peterson}, B.~M. 2006, \apj, 641, 689

\bibitem[{{Worseck} {et~al.}(2007){Worseck}, {Fechner}, {Wisotzki}, \&
  {Dall'Aglio}}]{worseck07:indirect.transverse.proximity}
{Worseck}, G., {Fechner}, C., {Wisotzki}, L., \& {Dall'Aglio}, A. 2007, \aap,
  473, 805

\bibitem[{{Worseck} \&
  {Wisotzki}(2006)}]{worseck06:indirect.transverse.proximity}
{Worseck}, G., \& {Wisotzki}, L. 2006, \aap, 450, 495

\bibitem[{{Wyithe} \& {Loeb}(2002)}]{wyitheloeb:sam}
{Wyithe}, J.~S.~B., \& {Loeb}, A. 2002, \apj, 581, 886

\bibitem[{{Younger} {et~al.}(2008){Younger}, {Hopkins}, {Cox}, \&
  {Hernquist}}]{younger:minor.mergers}
{Younger}, J.~D., {Hopkins}, P.~F., {Cox}, T.~J., \& {Hernquist}, L. 2008,
  \apj, 686, 815

\bibitem[{{Yu} \& {Lu}(2004)}]{yulu:bhmf}
{Yu}, Q., \& {Lu}, Y. 2004, \apj, 602, 603

\bibitem[{{Yu} \& {Lu}(2005)}]{yulu:stromgren}
---. 2005, \apj, 620, 31

\bibitem[{{Yu} \&
  {Lu}(2008)}]{yulu:lightcurve.constraints.from.bhmf.integration}
---. 2008, \apj, 689, 732

\bibitem[{{Yu} {et~al.}(2005){Yu}, {Lu}, \& {Kauffmann}}]{yu:mdot.dist}
{Yu}, Q., {Lu}, Y., \& {Kauffmann}, G. 2005, \apj, 634, 901

\bibitem[{{Yu} \& {Tremaine}(2002)}]{yutremaine:bhmf}
{Yu}, Q., \& {Tremaine}, S. 2002, \mnras, 335, 965

\bibitem[{{Yuan} \& {Narayan}(2004)}]{xbongs}
{Yuan}, F., \& {Narayan}, R. 2004, \apj, 612, 724

\bibitem[{{Zakamska} {et~al.}(2008){Zakamska}, {G{\'o}mez}, {Strauss}, \&
  {Krolik}}]{zakamska:mir.seds.type2.qso.transition.at.special.lum}
{Zakamska}, N.~L., {G{\'o}mez}, L., {Strauss}, M.~A., \& {Krolik}, J.~H. 2008,
  \aj, 136, 1607

\bibitem[{{Zakamska} {et~al.}(2006)}]{zakamska:qso.hosts}
{Zakamska}, N.~L., {et~al.} 2006, \aj, 132, 1496

\end{thebibliography}

\end{document}